\documentclass[showpacs,preprintnumbers,amsmath,amssymb,pra]{revtex4}
\usepackage{graphicx}
\usepackage{dcolumn}
\usepackage{bm}
\usepackage{times}
\usepackage[colorlinks,citecolor=blue,linkcolor=red]{hyperref}
\usepackage{color}

\begin{document}

\title{A polaron theory of quantum thermal transistor in nonequilibrium three-level systems}

\author{Chen Wang$^{1,}$}\email{wangchenyifang@gmail.com}
\author{Dazhi Xu$^{2,}$}\email{dzxu@bit.edu.cn}
\address{
$^{1}$Department of Physics, Zhejiang Normal University, Jinhua 321004, Zhejiang , P. R. China\\
$^{2}$School of Physics and Center for Quantum Technology Research,
Beijing Institute of Technology, 5 South Zhongguancun Street, Beijing 100081, China}

\date{\today}

\begin{abstract}
We investigate the quantum thermal transistor effect in nonequilibrium three-level systems by applying the polaron transformed Redfield equation combined with full counting statistics. The steady state heat currents are obtained via this unified approach over a wide region of system-bath coupling, and can be analytically reduced to the Redfield and nonequilibrium noninteracting blip approximation results in the weak and strong coupling limits, respectively. A giant heat amplification phenomenon emerges in the strong system-bath coupling limit,
where transitions mediated by the middle thermal bath is found to be crucial to  unravel the underlying mechanism.
Moreover, the heat amplification is also exhibited with moderate coupling strength, which can be properly explained within the polaron framework.
\end{abstract}

\maketitle


\section{Introduction}

According to the Clausius statement~\cite{rclausius1879book}, the heat flow from the hot source to the cold drain naturally occurs driven by the thermodynamic bias. Without violating this fundamental law of thermodynamics, great efforts have been paid to find other ways of conducting the heat flow~\cite{mesposito2015prl,gkatz2016entropy,gbenenti2017pr,dsegal2008prl,jren2010prl,kmicadei2019nc}.
Accompanying with the rapid progress in quantum technology, the control of heat flow becomes an increasingly important issue in quantum computation~\cite{lwang2007prl} and quantum measurement~\cite{lcui2017science,dsegal2017science}.

The thermal transistor, one of the novel phenomena in quantum thermal transport, which was initially proposed by B. Li and coworkers~\cite{bli2005apl,nbli2012rmp}.
In particular, heat amplification and negative differential thermal conductance (NDTC) are considered as two main components of the thermal transistor.
Heat amplification describes an effect within the three-terminal setup,
that the tiny modification of the base current will dramatically change the current
at the collector and emitter, which enables the efficient energy transport~\cite{bli2005apl}.
While the NDTC effect is characterized by the suppression of the heat flow with increase of temperature bias within the two-terminal setup~\cite{dhhe2009prb,dhhe2010pre,dhhe2014pre}.

Later, the spin-based fully quantum thermal transistor was  proposed by K. Joulain {\emph{et al.}}~\cite{kjoulain2016prl}, which is composed by three coupled-qubits, each interacting with one individual thermal bath.
The qubit-qubit interaction within the system is found to be crucial to exhibit the heat amplification.
Consequently, The importance of the system nonlinearity and long-ranging interaction on the transistor effect
has also been unraveled in various coupled-qubits systems~\cite{bqguo2018pre,bqguo2019pre,jydu2019pre}.
Simultaneously, the strong qubit-bath interaction is revealed to be another key factor to cause giant heat amplification
in the two and three qubits systems~\cite{cwang2018pra,hliu2019pre}, which also stems from the NDTC effect.
Hence, the NDTC is widely accepted as the crucial intergradient to realize the giant heat amplification.
However, J. H. Jiang \emph{et al.} pointed out that heat amplification can be realized via the inelastic transfer process
independent of NDTC~\cite{jhjiang2015prb}.
Hence, questions are raised that can these two types of heat amplification coexist in one quantum system?
Moreover, what is the minimal quantum system to show such heat amplification and NDTC?

Very recently, a three-level quantum heat transistor was preliminarily investigated by S. H. Su \emph{et al.},
which stressed the significant influence of quantum coherence in the heat amplification~\cite{shsu2018arxiv}.
However, the calculations are based on the phenomenological Lindblad equation,
which cannot be generalized to describe the heat transport beyond the weak system-bath coupling.
Considering the scientific importance and extensive application of the nonequilibrium three-level systems~\cite{hscovil1959prl,htquan2007pre,dzxu2016njp,tkrause2015jcp,eboukobza2007prl},
it is intriguing to give a comprehensive picture of the heat transistor behavior.
In particular, it is worthwhile to give a unified analysis on effect of the system-bath interaction from the weak to strong coupling regime on the heat amplification and NDTC.

\begin{figure}[tbp]
\begin{center}
\includegraphics[scale=0.6]{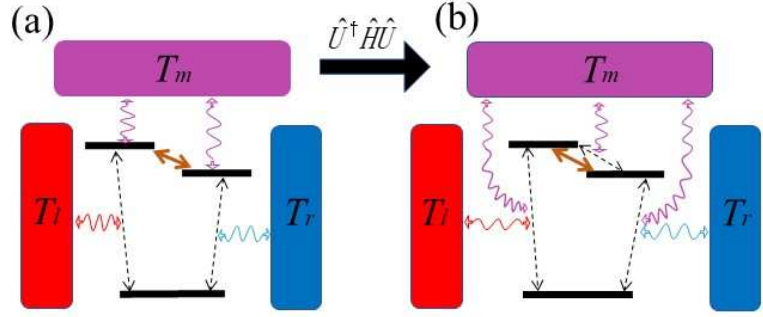}
\end{center}
\caption{(Color online) Schematics of the nonequilbrium V-type three-level system in
(a)the original framework described at Eq.~(\ref{h0}) and
(b)the polaron framework described at Eq.~(\ref{ct1}).
The three vertical solid black lines represent the central three-level model[$|u{\rangle}~(u=l,r,0)$], and the double-arrowed solid brown line shows the coherent tunneling between two excited states $|l{\rangle}$ and $|r{\rangle}$;
the rectangular left red, top-middle purple and right blue boxes describe three thermal baths, which are characterized by temperatures
$T_l$, $T_m$ and $T_r$, respectively;
the double-arrowed solid red, purple and blue curves describe interactions between the system and thermal baths,
and the double-arrowed dashed black lines describe transitions between different states assisted by phonons in the
corresponding thermal bath.
}~\label{fig0}
\end{figure}

In this paper, we devote to investigating quantum thermal transport in a quantum thermal transistor, which is composed by a three-level quantum system
within the three-terminal setup in Fig.~\ref{fig0}(a) by applying the polaron-transformed Redfield equation (PTRE), detailed in Sec. II.
In Sec. III, the heat currents are obtained from PTRE combined with full counting statistics (FCS)~\cite{levitov1992jetp,levitov1996jmp,mesposito2009rmp},
and are reduced to the Redfield and nonequilibrium noninteracting blip approximation (NIBA) schemes.
In Sec IV. A, the giant heat amplification is explored at strong system-middle bath interaction, and the underlying mechanism is proposed within the nonequilibrium NIBA.
While in Sec. IV B, another giant heat amplification and NDTC are found  at moderate system-middle bath coupling, which can not be explained by the Redfield equation.
Finally, we give a concise summary in Sec. V.

\section{Model and method}
We first introduce the nonequilibrium three-level system and the framework of polaron transformation.
Then, the polaron-transformed Redfield equation (PTRE) is applied to obtain the reduced system density matrix.

\subsection{Nonequilibrium three-level system}
The total Hamiltonian of the nonequilibrium three-level system interacting with three thermal baths, as shown in Fig.~\ref{fig0}(a), is described as
\begin{eqnarray}~\label{h0}
\hat{H}=\hat{H}_s+\sum_{u=l,r,m}(\hat{H}^u_b+\hat{V}_u).
\end{eqnarray}
The quantum three-level system is expressed as
\begin{eqnarray}
\hat{H}_s={\sum_{u=l,r,0}\varepsilon_u|u{\rangle}{\langle}u|}+\Delta(|l{\rangle}{\langle}r|+|r{\rangle}{\langle}l|),
\end{eqnarray}
where $|l(r){\rangle}$ is the left (right) excited state with the occupation energy $\varepsilon_{l(r)}$,
and $|0{\rangle}$ is the ground state with $\varepsilon_0=0$ for simplicity.
Specifically, for $\varepsilon_{l(r)}>0$, the three-level system corresponds to a V-type configuration.
Whereas for $\varepsilon_{l(r)}<0$, it corresponds to a $\Lambda$-type configuration with excited state $|0{\rangle}$ and ground states $|l{\rangle}$ and $|r{\rangle}$.
In the following, {our work is based on the V-type system}, without losing any generality.
The Hamiltonian of the $u$th~$(u=l,r,m)$ thermal bath is described as $\hat{H}^u_b=\sum_k\omega_k\hat{b}^\dag_{k,u}\hat{b}_{k,u}$,
where $\hat{b}^\dag_{k,u}~(\hat{b}_{k,u})$ creates (annihilates) one phonon in the $u$th bath with the frequency $\omega_k$.
The interaction between the left (right) excited state and the corresponding bath is given by
\begin{eqnarray}
\hat{V}_u=(\hat{S}^\dag_u+\hat{S}_u)\sum_k(g_{k,u}\hat{b}^\dag_{k,u}+g^*_{k,u}\hat{b}_{k,u}),\ u=l,r,
\end{eqnarray}
where $\hat{S}_u=|0{\rangle}{\langle}u|$, $\hat{S}^\dag_u=|u{\rangle}{\langle}0|$.
The quantum dissipation of the two excited states {induced by} the middle bath is modeled by a diagonal interaction~\cite{dzxu2016njp,shsu2018arxiv}, which reads
\begin{eqnarray}~\label{vm0}
\hat{V}_m={(|l{\rangle}{\langle}l|-|r{\rangle}{\langle}r|)\sum_{k}}(g_{k,m}\hat{b}^\dag_{k,m}+g^{*}_{k,m}\hat{b}_{k,m}).
\end{eqnarray}
The $u$th thermal bath here is characterized by the spectral function
$\Lambda_u(x)=4\pi\sum_k|g_{k,u}|^2\delta(x-\omega_k)$, where $g_{k,u}$ is the system-bath coupling strength. The spectral functions are assumed to be the super-Ohmic form
\begin{subequations}
\begin{align}
\Lambda_{l(r)}(x)=&\pi\gamma_{l(r)}\frac{x^3}{\omega^2_{c}}e^{-|x|/\omega_c},\\
\Lambda_{m}(x)=&\pi\alpha_{m}\frac{x^3}{\omega^2_{c}}e^{-|x|/\omega_c},
\label{spectrum}
\end{align}
\end{subequations}
where $\gamma_{l(r)}$ and $\alpha_m$ are the {system-bath coupling strengths corresponding to the $l(r)$th bath} and the $m$th bath, respectively. The cut-off frequency is $\omega_c$.
The super-Ohmic bath has been extensively included to investigate quantum dissipation~\cite{ajleggett1987rmp,sjang2013njp}
and quantum energy transport~\cite{anazir2009prl,dzxu2016fp,mqin2019pra}.

{In order to} consider the interaction between the excited states and the middle bath beyond the weak coupling limit,
we apply a canonical transformation [see Fig.~\ref{fig0}(b)]~\cite{sjang2013njp,cwang2015sr}
\begin{eqnarray}~\label{ct}
\hat{H}^\prime=\hat{U}^{\dag}\hat{H}\hat{U},
\end{eqnarray}
with $\hat{U}={\exp}[i\hat{B}(|l{\rangle}{\langle}l|-|r{\rangle}{\langle}r|)]$ and the collective phonon momentum operator $\hat{B}=i\sum_k(\frac{g_{k,m}}{\omega_k}\hat{b}^\dag_{k,{m}}-\textrm{H.c.})$.
{The resultant Hamiltonian is given by}
\begin{eqnarray}~\label{ct1}
\hat{H}^\prime=\hat{H}^\prime_s+\sum_{u=l,m,r}(\hat{H}^u_b+\hat{V}^\prime_u).
\end{eqnarray}
The modified system Hamiltonian is given by
\begin{eqnarray}~\label{hps}
\hat{H}^\prime_s=\overline{\varepsilon}\hat{N}
+\delta{\varepsilon}\hat{\sigma}_z+\eta\Delta\hat{\sigma}_x,
\end{eqnarray}
with the average occupation energy $\overline{\varepsilon}={(\varepsilon_l+\varepsilon_r)}/2-\sum_k|g_{k,m}|^2/\omega_k$,
the energy bias $\delta\varepsilon=(\varepsilon_l-\varepsilon_r)/2$,
the {excitation} number operator $\hat{N}=|l{\rangle}{\langle}l|+|r{\rangle}{\langle}r|$,
the bias operator $\hat{\sigma}_z=|l{\rangle}{\langle}l|-|r{\rangle}{\langle}r|$, {and} the tunneling operator
 $\hat{\sigma}_x=|l{\rangle}{\langle}r|+|r{\rangle}{\langle}l|$.
The renormalization factor $\eta={\langle}e^{{\pm}2i\hat{B}}{\rangle}$
is the expectation value of the bath displacement operator $e^{{\pm}2i\hat{B}}$ with respect to the middle bath equilibrium state.
It should be noted that $\hat{N}=\hat{I}-|0{\rangle}{\langle}0|$ with $\hat{I}$ the unit operator.
The transformed system Hamiltonian $\hat{H}^\prime_s$ can be exactly solved as
$\hat{H}^\prime_s|\pm{\rangle_\eta}=E_{\pm}|\pm{\rangle_\eta}$, where the eigenstates are
\begin{subequations}~\label{eigenstate}
\begin{align}
|+{\rangle}_{\eta}=&\cos\frac{\theta}{2}|l{\rangle}+\sin\frac{\theta}{2}|r{\rangle},\\~\label{eigenstate12}
|-{\rangle}_{\eta}=&-\sin\frac{\theta}{2}|l{\rangle}+\cos\frac{\theta}{2}|r{\rangle},
\end{align}
\end{subequations}
with $\tan\theta=\eta\Delta/\delta{\varepsilon}$
and the eigenvalues $E_{\pm}=\overline{\varepsilon}{\pm}\sqrt{(\delta\varepsilon)^2+(\eta\Delta)^2}$.

Moreover, the modified system-bath interaction is given by
\begin{subequations}
\begin{align}
\hat{V}^\prime_m=&{\Delta[\cos(2\hat{B})-\eta]\hat{\sigma}_x+\Delta\sin(2\hat{B})\hat{\sigma}_y},~\label{vprime1}\\
\hat{V}^\prime_u=&(e^{-i\hat{B}_u}{\hat{S}_u^\dag}+e^{i\hat{B}_u}\hat{S}_u)\sum_k(g_{k,u}\hat{b}^\dag_{k,u}+g^*_{k,u}\hat{b}_{k,u})~\label{vprime2},\ u=l,r,
\end{align}
\end{subequations}
with $\hat{\sigma}_y={-i}(|l{\rangle}{\langle}r|-|r{\rangle}{\langle}l|)$,
$\hat{B}_l=\hat{B}$ and $\hat{B}_r=-\hat{B}$.
All the interaction terms $\hat{V}^\prime_u~(u=l,m,r)$ of Eq.~(\ref{vprime1}) and Eq.~(\ref{vprime2}) imply multi-phonon transferring processes.
Specifically, $\hat{V}^\prime_m$ involves multiple phonons absorption or emission accompanying the transition between the two excited states,
which can be understood by {the expansion}
$\cos(2\hat{B})=\sum_{n=0}\frac{(2\hat{B})^{2n}}{(2n)!}$  and $\sin(2\hat{B})=\sum_{n=0}\frac{(2\hat{B})^{2n+1}}{(2n+1)!}$.
While the interaction between the left (right) bath phonon and the three-level system $\hat{V}^\prime_{u}$ now involves the polaron effect embodied in the displacement operator $\exp(\pm i\hat{B}_u)$ of the middle bath phonon modes.


\subsection{Polaron transformed master equation}
{It can be easily verified that the thermal average of the modified interaction ${\langle}\hat{V}^\prime_m{\rangle}$ is zero, which makes ${\langle}\hat{V}^\prime_m{\rangle}$ a properly perturbative term~\cite{cwang2015sr}. Therefore, we can apply the quantum master equation in the polaron frame to study the dynamics of the three-level system.}
Moreover, we assume the interaction between the system and the left (right) bath $\hat{V}^\prime_{l(r)}$ is weak.
Thus we can separately apply the perturbation theory with respect to two terms in Eq.~(\ref{vprime1}) and Eq.~(\ref{vprime2}).
{Accordingly, the PTRE based on the Born-Markov approximation can be written as}
\begin{eqnarray}
\frac{d}{dt}\hat{\rho}_s=-i[\hat{H}^\prime_s,\hat{\rho}_s]+\sum_{u=l,m,r}\mathcal{L}_u[\hat{\rho}_s],
\end{eqnarray}
where $\hat{\rho}_s$ is the density operator of the three-level system. The $m$th dissipator is specified as
\begin{eqnarray}
\mathcal{L}_m[\hat{\rho}_s]=\sum_{\alpha=x,y;\omega,\omega^\prime}
\gamma_\alpha(\omega^\prime)[\hat{P}_\alpha(\omega^\prime)\hat{\rho}_s\hat{P}_\alpha(\omega)
-\hat{P}_\alpha(\omega)\hat{P}_\alpha(\omega^\prime)\hat{\rho}_s]+\textrm{H.c.},
\end{eqnarray}
where the dissipation rates between two excited eigenstates are
\begin{subequations}
\begin{align}
\gamma_x(\omega)=&\eta^2\Delta^2\int^\infty_0d{\tau}e^{i\omega\tau}[\cosh\phi_m(\tau)-1],~\label{gammaxy1}\\
\gamma_y(\omega)=&\eta^2\Delta^2\int^\infty_0d{\tau}e^{i\omega\tau}\sinh\phi_m(\tau),~\label{gammaxy2}
\end{align}
\end{subequations}
with the correlation phase
\begin{eqnarray}
\phi_m(\tau)&=&4\sum_k\big{|}\frac{{g_{k,m}}}{\omega_k}\big{|}^2\{\cos(\omega_k\tau){[2n_{m}(\omega_k)+1]}-i\sin(\omega_k\tau)\}.
\end{eqnarray}
{The} operators $\hat{P}_\alpha(\omega)~(\alpha=x,y)$ {are the projective operators of the system eigenbasis, which are defined by}
$\hat{\sigma}_\alpha(-\tau)=\sum_{\omega}\hat{P}_\alpha(\omega)e^{i\omega\tau}$ with $\hat{P}_\alpha(-\omega)=\hat{P}^\dag_\alpha(\omega)$.
The rate $\gamma_y(\omega)$ describes the transition between the two excited eigenstates $|\pm{\rangle}_\eta$
involving odd number of phonons from the middle thermal bath. The bath average phonon number is $n_u(\omega)=1/[\exp(\omega/T_u)-1],\ u=r,l,m$, with $T_u$ the temperature of the $u$th bath.
The approximated expression of the real part of $\gamma_{y}(\omega)$ to the first-order of $\phi_m$ reads
$\textrm{Re}{[\gamma_{y}(\omega)]\approx 4\pi\eta^2\Delta^2\sum_k|\frac{{g_{k,m}}}{\omega_k}|^2[n_{m}(\omega_k)+1]\delta(\omega-\omega_k)}$, which contains the sequential process
of creating one phonon {with frequency $\omega=\omega_k$} in the $m$th bath.
{A direct consequence of the polaron transformation is that the dissipative rates $\gamma_x(\omega)$ and $\gamma_y(\omega)$ contain all the high-order terms of $\phi_m$, which can be understood as the contribution of the multiple-phonon correlation. In the strong system-bath coupling strength regime, such high-order correlations should be properly incorporated in the evolution of the open quantum system.}

The dissipators associated with the left and right bath are given by
\begin{eqnarray}
\mathcal{L}_u[\hat{\rho}_s]&=&{\sum_{\omega,\omega^\prime}}[\kappa_{u,-}(\omega^\prime)\hat{Q}_u(\omega^\prime)\hat{\rho}_s\hat{Q}^{\dag}_u(\omega)
+\kappa_{u,+}(\omega^\prime)\hat{Q}^\dag_u(\omega^\prime)\hat{\rho}_s\hat{Q}_u(\omega)\nonumber\\
&&-\kappa_{u,+}(\omega^\prime)\hat{Q}_u(\omega)\hat{Q}^\dag_u(\omega^\prime)\hat{\rho}_s
-\kappa_{u,-}(\omega^\prime)\hat{Q}^\dag_u(\omega)\hat{Q}_u(\omega^\prime)\hat{\rho}_s]+\textrm{H.c.},
\end{eqnarray}
where the system part operators are defined by
$\hat{S}_u(-\tau)=\sum_{\omega}\hat{Q}_u(\omega)e^{i\omega\tau}$,
and the dissipation rates are
\begin{subequations}
\begin{align}
\kappa_{u,+}(\omega)=&\int^\infty_{-\infty}\frac{d{\omega_1}}{4\pi}{\Lambda_u}(\omega_1)n_u(\omega_1)C_u(\omega_1-\omega),~\label{Gammau1}\\
\kappa_{u,-}(\omega)=&\int^\infty_{-\infty}\frac{d{\omega_1}}{4\pi}{\Lambda_u}(\omega_1)[1+n_u(\omega_1)]C_u(-\omega_1+\omega).~\label{Gammau2}
\end{align}
\end{subequations}
The phonon correlation function above is defined by
\begin{eqnarray}
C_u(\omega)={\eta^2_u}\int^\infty_0d{\tau}e^{i\omega\tau}e^{\phi_m(\tau)/4},
\end{eqnarray}
with {$\eta_l=\langle e^{i\hat{B}}\rangle$} and $\eta_r=\langle e^{-i\hat{B}}\rangle$. The transition rates in Eq.~(\ref{Gammau1}) and Eq.~(\ref{Gammau2}) demonstrate the joint contribution of the left (right) and the middle thermal baths on the nonequilibrium energy exchange.
Specifically, $\kappa_{l(r),+}(\omega)$ describes the process that one phonon with the frequency $\omega_1$ is emitted from the $l(r)$th thermal bath to assist the excitation from $|0{\rangle}$ to the eigenstate with energy gap $\omega$, and
the resultant energy $\omega_1-\omega>0~(\omega_1-\omega<0)$ is released (absorbed) into (from) the $m$th bath.
While the rate $\kappa_{l(r),-}(\omega)$ shows the transition {that} one phonon with frequency $\omega_1$ is absorbed by the $l(r)$th thermal bath,
and the three-level system is relaxed from the excited state with energy $\omega$ into $|0{\rangle}$.

\section{Steady state heat currents}

In this section, we first briefly introduce the concept of { full counting statistics (FCS)}.
Then we generalize the PTRE to incorporate the counting parameters, by means of which we derive the steady state heat currents.
Finally, we obtain the analytical expression of heat currents in both the strong and weak coupling limits via the nonequilibrium NIBA approach and the Redfield equation, respectively.

\subsection{Brief introduction of the FCS}
Full counting statistics~\cite{mesposito2009rmp}, also termed as the large deviation theory~\cite{htouchette2009pr},
is considered as a typical approach to study {the energy or particle} flow and its fluctuations~\cite{jren2010prl,jcerrillo2016prb,qshi2017prb,dsegal2018njp},
which is based on the two-time measurement protocol.
Generally, within the time interval $[0,\tau]$ the characteristic function to demonstrate the thermal transport into the $u$th thermal bath is expressed as
\begin{eqnarray}~\label{zv}
\mathcal{Z}(\chi_u,\tau)=\textrm{Tr}[e^{i\chi_u\hat{H}^u_b(0)}e^{-i\chi_u\hat{H}^u_b(\tau)}\hat{\rho}_{\textrm{tot}}(0)],
\end{eqnarray}
where $\chi_u$ is the {parameter counting the heat flow from the $u$th bath},
$\hat{H}^u_b(\tau)=e^{i\hat{H}\tau}\hat{H}^u_be^{-i\hat{H}\tau}$ {is the bath Hamiltonian in the Heisenberg picture},
$\hat{H}$ is the total Hamiltonian including system and baths,
and $\hat{\rho}_{\textrm{tot}}(0)$ is the initial state density operator of the total system.
Moreover, by assuming the commutating condition $[\hat{\rho}_{\textrm{tot}}(0),\hat{H}^u_b]=0$,
the characteristic function of Eq.~(\ref{zv}) can be re-expressed as~\cite{dsegal2018njp}
\begin{eqnarray}
\mathcal{Z}(\chi_u,\tau)&=&\textrm{Tr}[\hat{M}_{-\chi_u}(\tau)\hat{\rho}_{\textrm{tot}}(0)\hat{M}^{\dag}_{\chi_u}(\tau)]\nonumber\\
&\equiv&\textrm{Tr}[\hat{\rho}_{\chi_u}(\tau)]~\label{Z},
\end{eqnarray}
where the evolution operator is  $\hat{M}_{\chi_u}(\tau)=e^{-i\hat{H}_{\chi_u}\tau}$ with the counting parameter dependent Hamiltonian
$\hat{H}_{\chi_u}=e^{i\chi_u\hat{H}^u_b/2}\hat{H}e^{-i\chi_u\hat{H}^u_b/2}$.
Therefore, via the cumulant-generating function
$\mathcal{F}(\chi_u)=\lim_{\tau{\rightarrow}\infty}\frac{1}{\tau}\ln\mathcal{Z}(\chi_u,\tau)$,
the steady state heat flux can be straightforwardly  obtained as
\begin{eqnarray}
J_u=\frac{{\partial}\mathcal{F}(\chi_u)}{{\partial}(i\chi_u)}\bigg{|}_{\chi_u=0}.
\end{eqnarray}

\subsection{Generalized quantum master equation}

It can be seen from the above subsection, the key of accessing $J_u$ via FCS lies in solving the counting parameter dressed density operator $\hat{\rho}_{\chi_u}(\tau)$~\cite{mesposito2009rmp}. The time evolution of $\hat{\rho}_{\chi_u}(\tau)$ has been explicitly given by Eq.~(\ref{Z}), which can be equivalently written in the differential form of the quantum Liouvillian equation with an effective Hamiltonian including the counting parameters
\begin{eqnarray}
{\hat{H}_{\{\chi\}}}=\hat{H}_s+\sum_{u=l,m,r}\hat{H}^u_b+\hat{V}_m+\sum_{u=l,r}\hat{V}_u(\chi_u),
\end{eqnarray}
where $\{\chi\}=(\chi_l,\chi_r)$ {is a set of parameters counting both the heat flows from the left and right baths. The} modified system-bath interactions are
\begin{eqnarray}~\label{vuchiu}
\hat{V}_u(\chi_u)=(\hat{S}^{\dag}_u+\hat{S}_u)\sum_k(g_{k,u}e^{i\omega_k\chi_u/2}\hat{b}^\dag_{k,u}+\textrm{H.c.}),\ u=l,r.
\end{eqnarray}
By the analogous unitary transformation of Eq.~(\ref{ct}), $\hat{V}_u(\chi_u)$ is transformed to
\begin{eqnarray}
\hat{V}^\prime_u(\chi_u)=(e^{-i\hat{B}_u}\hat{S}^{\dag}_u+e^{i\hat{B}_u}\hat{S}_u)\sum_k(g_{k,u}e^{i\omega_k\chi_u/2}\hat{b}^\dag_{k,u}+\textrm{H.c.}).
\end{eqnarray}
With the same approach introduced in the Sec II.B, we obtain the {PTRE} of system density operator $\hat{\rho}_{\{\chi\}}$, which has been marked by the counting parameters as
\begin{eqnarray}~\label{dechi1}
\frac{d}{dt}\hat{\rho}_{\{\chi\}}=-i[\hat{H}^\prime_s,\hat{\rho}_{\{\chi\}}]+\mathcal{L}_m[\hat{\rho}_{\{\chi\}}]
+\sum_{u=l,r}\mathcal{L}^u_{\chi_u}[\hat{\rho}_{\{\chi\}}].
\end{eqnarray}
Here, the generalized dissipator is
\begin{eqnarray}
\mathcal{L}^u_{\chi_u}[\hat{\rho}_{\{\chi\}}]&=&
{\sum_{\omega,\omega^\prime}}\{\kappa_{u,+}(\omega^\prime,\chi_u)\hat{Q}^\dag_u(\omega^\prime)\hat{\rho}_{\{\chi\}}\hat{Q}_u(\omega)
+\kappa_{u,-}(\omega^\prime,\chi_u)\hat{Q}_u(\omega^\prime)\hat{\rho}_{\{\chi\}}\hat{Q}^\dag_u(\omega)\\
&&+\kappa^{*}_{u,+}(\omega^\prime,-\chi_u)\hat{Q}^\dag_u(\omega)\hat{\rho}_{\{\chi\}}\hat{Q}_u(\omega^\prime)
+\kappa^{*}_{u,-}(\omega^\prime,-\chi_u)\hat{Q}_u(\omega)\hat{\rho}_{\{\chi\}}\hat{Q}^\dag_u(\omega^\prime)\nonumber\\
&&-[\kappa_{u,+}(\omega^\prime)\hat{Q}_u(\omega)\hat{Q}^\dag_u(\omega^\prime)\hat{\rho}_s
+\kappa_{u,-}(\omega^\prime)\hat{Q}^\dag_u(\omega)\hat{Q}_u(\omega^\prime)\hat{\rho}_s+\textrm{H.c.}]\},\nonumber
\end{eqnarray}
with the generalized dissipation rates
\begin{subequations}
\begin{align}
\kappa_{u,+}(\omega^\prime,\chi_u)=&\int^{\infty}_{-\infty}\frac{{d\omega_1}}{4\pi}{\Lambda}(\omega_1)n_u(\omega_1)
e^{-i\omega_1\chi_u}C_{u}(\omega_1-\omega^\prime),~\label{kappa1}\\
\kappa_{u,-}(\omega^\prime,\chi_u)=&\int^{\infty}_{-\infty}\frac{{d\omega_1}}{4\pi}{\Lambda}(\omega_1)[1+n_u(\omega_1)]
e^{i\omega_1\chi_u}C_{u}(-\omega_1+\omega^\prime)~\label{kappa2}.
\end{align}
\end{subequations}
In absence of the counting parameters($\chi_l=0,\chi_r=0$), the density operator $\hat{\rho}_{\{\chi\}}$, the dissipator $\mathcal{L}^u_{\chi_u}[\hat{\rho}_{\{\chi\}}]$ and dissipation rates $\kappa_{u,\pm}(\omega^\prime,\chi_u)$ are reduced to the original $\hat{\rho}_s$, $\mathcal{L}_u[\hat{\rho}_{s}]$
and $\kappa_{u,\pm}(\omega^\prime)$ defined in the Sec.IIB, respectively.

Furthermore, we re-express the dynamical equation of Eq.~(\ref{dechi1}) by
\begin{eqnarray}~\label{LE}
\frac{d}{dt}|\mathbf{P}{\rangle\rangle}=\mathbb{L}_{\{\chi\}}|\mathbf{P}{\rangle\rangle},~\label{LE}
\end{eqnarray}
where $|\mathbf{P}{\rangle\rangle}=[\rho_{++},\rho_{--},\rho_{00},\rho_{+-},\rho_{-+}]^T$ is the vector form of the reduced density matrix,
with $\rho_{ij}={\langle}i|\hat{\rho}_s|j{\rangle}~(i,j=0,\pm)$,
and $\mathbb{L}_{\{\chi\}}$ is the super-operator defined according to Eq.~(\ref{dechi1}).
{The off-diagonal terms $\rho_{0\pm}$ are decoupled from Eq.~(\ref{LE}) and is irrelevant with the following discussion.} Therefore, the heat currents flow into the left and right thermal baths can be expressed as
\begin{eqnarray}
J_{u}={\langle\langle}\mathbf{I}|\frac{{\partial}\mathbb{L}_{\{\chi\}}}{{\partial}(i\chi_{u})}\bigg{|}_{{\chi}=0}|\mathbf{P}{\rangle}{\rangle},
\end{eqnarray}
with ${\langle\langle}\mathbf{I}|=[1,1,1,0,0]$ and $u=l,r$.
The steady state current into the middle bath is obtained by the energy conservation condition as
$J_{m}=-J_{l}-J_{r}$.

\begin{figure}[tbp]
\begin{center}
\includegraphics[scale=0.6]{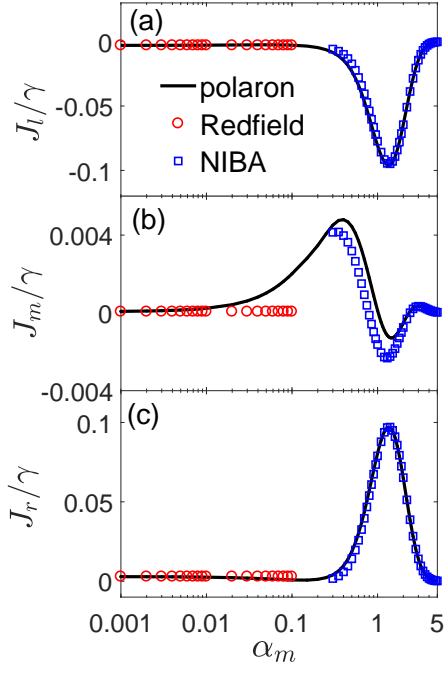}
\end{center}
\caption{(Color online) Steady state heat currents (a) $J_l/\gamma$, (b) $J_m/\gamma$, and (c) $J_r/\gamma$ as a function of the coupling strength $\alpha_m$,
with $\gamma=0.0002$.
The red circles is based on the Redfield scheme, the blue squares is based on the nonequilibrium noninteracting blip approximation (NIBA),
and black solid line is calculated from the nonequilibrium polaron-transformed Redfield approach.
The other parameters are given by
$\varepsilon_l=1.0$, $\varepsilon_r=0.6$, $\Delta=0.6$, $\omega_c=10$, $T_l=2$, $T_m=1.2$, and $T_r=0.4$.
}~\label{fig1}
\end{figure}

These three steady state currents are exhibited in Fig.~\ref{fig1}. It is found that the currents are all significantly enhanced in the moderate coupling regime around $\alpha_m{\in}(0.5,2)$, but are suppressed in both the strong and weak coupling limits. Moreover, the current flows into the middle bath is more sensitive in response to the change of $\alpha_m$. It can be seen that $J_m$ begins to increase significantly when $\alpha_m$ is around 0.01, which is one order smaller than those of the other two currents.

The PTRE combined with FCS has been successfully introduced to investigate quantum thermal transport in the nonequilibrium spin-boson systems~\cite{cwang2015sr,dzxu2016fp,cwang2017pra},
which is able to fully bridge the strong and weak system-bath coupling limits. In the following, we extend such {method to} the nonequilibrium three-level system.

\begin{figure}[tbp]
\begin{center}
\includegraphics[scale=0.5]{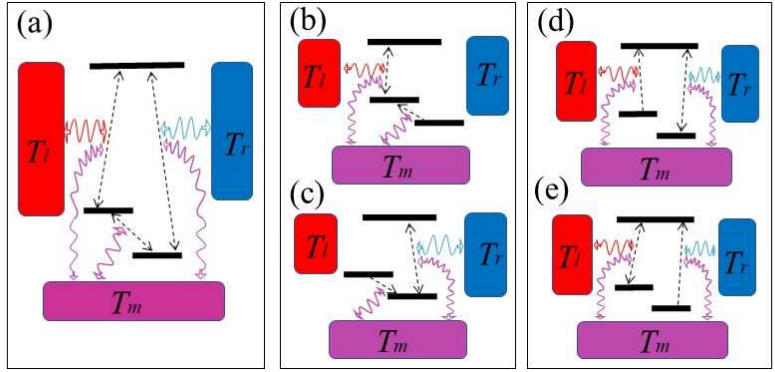}
\end{center}
\caption{(Color online)
(a) The globally cyclic transitions contributed by the $G^{\pm}_lG^{\pm}_mG^{\mp}_r/\mathcal{A}$,
and the  locally conditional transitions contributed by
(b) $G^-_mG^+_lG^-_l/\mathcal{A}$, (c)$G^+_mG^+_rG^-_r/\mathcal{A}$,
(d) $G^-_rG^+_lG^-_l/\mathcal{A}$ and
(e) $G^-_lG^+_rG^-_r/\mathcal{A}$ within the nonequilibrium NIBA scheme, respectively.
The vertical solid black line at top represents $|0{\rangle}$,
two vertical  solid black lines at bottom describe renormalized energy levels $|l(r){\rangle}_{\eta}$ with the energies given by Eq.~(\ref{eu}),
and the other symbols are the same as shown in Fig.~\ref{fig0}.
}~\label{fig1pp}
\end{figure}

\subsection{Currents with limiting couplings}
Generally, it is rather difficult to analytically obtain the expression of steady state heat currents with arbitrary coupling strength $\alpha_m$
between two excited states and middle thermal bath. Fortunately, in the strong and weak coupling limits, the NIBA approach and Redfield equation can adequately describe the steady state of the system, respectively. The validity of these two approaches can be checked numerically compared with the results of PTRE, meanwhile they are more simple in the form than PTRE and lead to the analytical expressions of the currents.

\subsubsection{Nonequilibrium NIBA}
In the strong coupling limit, the renormalization factors are dramatically suppressed,
i.e. $\eta{\ll}1$.
The eigenstates are reduced to the localized {ones} $|+{\rangle}_\eta{\approx}|l{\rangle}$,
$|-{\rangle}_\eta{\approx}|r{\rangle}$,
and the renormalized energies generally become negative
\begin{eqnarray}~\label{eu}
E_u=(\varepsilon_u-\sum_{k}\frac{|g_{k,m}|^2}{\omega_k})<0,\ u=l,r.
\end{eqnarray}
In this case, the configuration of the three-level system becomes the $\Lambda$-type as shown in Fig.~\ref{fig1pp}(a). Hence, the nonequilibrium PTRE can be analytically solved (the details are given in Appendix A) and gives the heat currents explicitly as
\begin{eqnarray}~\label{jl0}
J_{l,\textrm{NIBA}}&=&\frac{1}{\mathcal{A}}[(G^+_lG^+_mG^-_r{\langle}\omega{\rangle}_{l,+}-G^-_lG^-_mG^+_r{\langle}\omega{\rangle}_{l,-})\nonumber\\
&&+{(G^-_m+G^-_r)G^-_lG^+_l({\langle}\omega{\rangle}_{l,+}-{\langle}\omega{\rangle}_{l,-})]},
\end{eqnarray}
\begin{eqnarray}~\label{jr0}
J_{r,\textrm{NIBA}}&=&\frac{1}{\mathcal{A}}[(G^-_lG^-_mG^+_r{\langle}\omega{\rangle}_{r,+}-G^+_lG^+_mG^-_r{\langle}\omega{\rangle}_{r,-})\nonumber\\
&&+{(G^+_m+G^-_l)G^+_rG^-_r({\langle}\omega{\rangle}_{r,+}-{\langle}\omega{\rangle}_{r,-})]},
\end{eqnarray}
and $J_m=-J_l-J_r$, where the coefficient is
$\mathcal{A}=(G^+_m+G^-_m)(G^+_l+G^+_r)+G^+_mG^-_r+G^-_mG^-_l+G^-_lG^+_r+G^-_r(G^+_l+G^-_l)$,
the transition rates are
\begin{subequations}
\begin{align}
G^{\pm}_m=&\int^\infty_{-\infty}d{\tau}e^{{\pm}i(\varepsilon_l-\varepsilon_r)\tau}\eta^2{e^{\phi_m(\tau)}}~\label{gm},\\
G_{u}^{+}=&\frac{1}{4\pi}\int^\infty_{-\infty}{d\omega_1}\Lambda_u(\omega_1)[1+n_u(\omega_1)]
[{C}_u(-E_u-\omega_1)+H.c.]~\label{gup},\\
G_{u}^{-}=&\frac{1}{4\pi}\int^\infty_{-\infty}{d\omega_1}\Lambda_u(\omega_1)n_u(\omega_1)
[{C}_u(\omega_1+E_u)+H.c.]~\label{gum},
\end{align}
 \end{subequations}
and the average energy quanta into the $u$th thermal bath are
\begin{subequations}
\begin{align}
{\langle}\omega{\rangle}_{u,+}=&\frac{1}{4{\pi}G^+_u}\int^\infty_{-\infty}{d\omega_1}\omega_1\Lambda_u(\omega_1)[1+n_u(\omega_1)]
[{C}_u(-E_u-\omega_1)+H.c.]~\label{oup},\\
{\langle}\omega{\rangle}_{u,-}=&\frac{1}{4{\pi}G^-_u}\int^\infty_{-\infty}{d\omega_1}\omega_1\Lambda_u(\omega_1)n_u(\omega_1)
[{C}_u(\omega_1+E_u)+H.c.]~\label{oum},
\end{align}
\end{subequations}
In the following, the subscript $u$ only represents $l$ or $r$ without further declaration.
The rates $G^{\pm}_{u}$ at Eq.~(\ref{gup}) and Eq.~(\ref{gum}) are contributed by two physical processes.
Take $G^{+}_u$ {for example},
(i) resonant energy relaxation from the state $|0{\rangle}$ to $|u{\rangle}$,
with the energy $E_u$ absorbed by the $u$th thermal bath;
(ii) off-resonant transport process, where thermal baths show non-additive cooperation.
As the three-level system release energy $E_u$,  part of the heat $\omega_1$ is absorbed by the $u$th bath,
whereas the left energy $(-E_u-\omega_1)$ is consumed by the middle bath.
Similarly, the rate $G^{-}_u$ describes the reversed process of $G^{+}_u$.

The currents $J_l$ and $J_r$ in Eq.~(\ref{jl0}) and Eq.~(\ref{jr0}) are contributed by three distinct types of thermal transport processes:
(i) globally cyclic transition in Fig.~\ref{fig1pp}(a), which is contributed by the cooperative rate
${G^{\pm}_lG^{\pm}_mG^{\mp}_r}/\mathcal{A}$
to carry the average energy quanta ${\langle}\omega{\rangle}_{u,+}\ ({\langle}\omega{\rangle}_{u,-})$ to (from) the $u$th bath.
(ii) local transition $|u{\rangle}{\leftrightarrow}|0{\rangle}$ mediated by the middle bath dependent rate $G^{\pm}_m$,
which transfers the energy quanta ${\langle}\omega{\rangle}_{u,+}-{\langle}\omega{\rangle}_{u,-}$.
Fig.~\ref{fig1pp}(b) and  Fig.~\ref{fig1pp}(c) illustrate these transition processes characterized by the rates $G^-_mG^+_lG^-_l/\mathcal{A}$ and $G^+_mG^+_rG^-_r/\mathcal{A}$, respectively.
(iii) local transition $|u{\rangle}{\leftrightarrow}|0{\rangle}$ mediated by the $u$th bath dependent rate $G^{-}_{u}$,
which is characterized by the rates $G^-_rG^+_lG^-_l/\mathcal{A}$ and
$G^-_lG^+_rG^-_r/\mathcal{A}$ as illustrated in Fig.~\ref{fig1pp}(d) and Fig.~\ref{fig1pp}(e), respectively.

We compare the heat currents calculated by the nonequilibrium NIBA with the ones calculated by the PTRE in Fig.~\ref{fig1}.
It is found that $J_{u}$ obtained by these two methods are consistent with each other in a wide regime of the coupling strength $\alpha_m$.
While $J_{m}$ obtained from the nonequilibrium NIBA shows apparently disparity with the result of the PTRE, unless the coupling strength increases to the regime $\alpha_m{\gtrsim}2$.

\subsubsection{Redfield scheme}
In the weak coupling limit, the renormalization factors become $\eta{\approx}1$.
The counting parameter dependent transition rates defined in Eq.~(\ref{kappa1}) and Eq.~(\ref{kappa2}) are simplified to
\begin{subequations}
\begin{align}
\textrm{Re}[\kappa_{u,+}(\omega^\prime,\chi_u)]{\approx}&\frac{1}{4}\Lambda(\omega^\prime)n_u(\omega^\prime)e^{-i\omega^\prime\chi_u}~\label{rup},\\
\textrm{Re}[\kappa_{u,-}(\omega^\prime,\chi_u)]{\approx}&\frac{1}{4}\Lambda(\omega^\prime)[1+n_u(\omega^\prime)]e^{i\omega^\prime\chi_u}~\label{rum}.
\end{align}
\end{subequations}
And the transition rates in Eq.~(\ref{gammaxy1}) and Eq.~(\ref{gammaxy2}) are reduced to
$\textrm{Re}[\gamma_x(\omega)]{\approx}0$,
and
$\textrm{Re}[\gamma_y(\omega)]{\approx}\frac{\Delta^2}{\omega^2}\Lambda_m(\omega)[1+n_m(\omega)]$,
where we only keep the lowest order terms of the correlation phase $\phi_m(\tau)$.
Then, the NE-PTRE in Eq.~(\ref{dechi1}) is reduced to the seminal Redfield equation(see Appendix B).
Consequently, the steady state currents are obtained as
\begin{eqnarray}~\label{jl}
J_l&=&\sum_{\xi=\pm}\frac{(1+\xi\cos\theta)}{4\mathcal{B}}E_{\xi}(\Gamma^e_++\Gamma^e_-)\{\kappa^a_{l,\xi}[\Gamma^e_{\xi}\Gamma^a_{\bar{\xi}}+(\Gamma^e_++\Gamma^e_-)\Gamma^{\xi}_p]
\nonumber\\
&&-\kappa^e_{l,\xi}(\Gamma^a_+\Gamma^a_-+\Gamma^a_+\Gamma^+_p+\Gamma^a_-\Gamma^-_p)\},\\
J_r&=&\sum_{\xi=\pm}\frac{(1-\xi\cos\theta)}{4\mathcal{B}}E_{\xi}(\Gamma^e_++\Gamma^e_-)\{\kappa^a_{r,\xi}[\Gamma^e_{\xi}\Gamma^a_{\bar{\xi}}+(\Gamma^e_++\Gamma^e_-)\Gamma^{\xi}_p]\nonumber\\
&&-\kappa^e_{r,\xi}(\Gamma^a_+\Gamma^a_-+\Gamma^a_+\Gamma^+_p+\Gamma^a_-\Gamma^-_p)\},
\end{eqnarray}
{where $\bar\xi\equiv-\xi$}. The current into the middle bath is
\begin{eqnarray}~\label{jm}
J_m=-(E_+-E_-)\frac{\Gamma^e_++\Gamma^e_-}{\mathcal{B}}(\Gamma^a_+\Gamma^e_-\Gamma^+_p-\Gamma^a_-\Gamma^e_+\Gamma^-_p),
\end{eqnarray}
where the coefficient is
$\mathcal{B}=\sum_{\xi=\pm}(\Gamma^a_\xi+\Gamma^e_++\Gamma^e_-)[\Gamma^e_\xi\Gamma^a_{\overline{\xi}}+(\Gamma^e_++\Gamma^e_-)\Gamma^\xi_p]$,
the combined rates are
\begin{subequations}
\begin{align}
\Gamma^{e(a)}_+=&\frac{1}{2}(\kappa^{e(a)}_{l,+}\cos^2\frac{\theta}{2}+\kappa^{e(a)}_{r,+}\sin^2\frac{\theta}{2}),~\label{gep}\\
\Gamma^{e(a)}_-=&\frac{1}{2}(\kappa^{e(a)}_{l,-}\sin^2\frac{\theta}{2}+\kappa^{e(a)}_{r,-}\cos^2\frac{\theta}{2}),~\label{gem}\\
\Gamma^{+(-)}_p=&\frac{\sin^2\theta}{2}\kappa^{e(a)}_p,~\label{gp}
\end{align}
\end{subequations}
and the local rates are
$\kappa^e_{u,\pm}=\Lambda_u(E_{\pm})n_u(E_{\pm})$,
$\kappa^a_{u,\pm}=\Lambda_u(E_{\pm})[1+n_u(E_{\pm})]$,
$\kappa^e_{p}=\Lambda_m(E_+-E_-)n_m(E_+-E_-)$,
and
$\kappa^a_{p}=\Lambda_m(E_+-E_-)[1+n_m(E_+-E_-)]$.

We plot the currents [Eqs.~(\ref{jl}-\ref{jm})] in Fig.~\ref{fig1} to analyze the valid regime of $\alpha_m$ by comparing with the counterpart based on the PTRE.
It is found that for $J_l$ and $J_r$ the Redfield scheme is applicable even for the system-middle bath coupling strength $\alpha{=}0.1$.
While for $J_m$, the Redfield scheme becomes invalid as system-middle bath coupling strength surpasses $0.01$, where the approximation of the transition rates in Eq.~(\ref{rup}) and Eq.~(\ref{rum}) break down.
This fact indicates that the influence of the phonons in the middle bath should be necessarily included to describe the transitions
between $|{\pm}{\rangle}_{\eta}$ and $|0{\rangle}$, which may enhance the energy flow into the middle bath accordingly.

In the following, based on the consistent analysis of the heat currents (particular for $J_m$) we approximately classify the strength of the system-middle bath interaction into three regimes:
(i) weak coupling regime $\alpha_m{<}0.01$; (ii) moderate coupling regime {$0.01\leq\alpha_m\leq2$};
(iii) strong coupling regime $\alpha_m>2$.

\section{Results and discussions}
Heat amplification and negative differential thermal conductance are considered as two crucial components of the quantum thermal transistor.
Particularly for heat amplification,
the transistor of the three terminal setup, the schematics of which are shown in Fig.~\ref{fig0}(a),
has the ability to significantly enhance the heat flow { the left or right terminal by a tiny modulation of the middle terminal {temperature}.
Formally, the amplification factor is defined as~\cite{bli2005apl}
\begin{eqnarray}
\beta_{u}=\big{|}{\partial}J_u/{\partial}J_m\big{|},\ u=l,r.
\end{eqnarray}
Moreover, owning to the flux conservation of the three-level system
$J_l+J_m+J_r=0$, the amplification factors $\beta_l$ and $\beta_r$
are related with
$\beta_l=\big{|}\beta_r+(-1)^\theta\big{|}$,
with $\theta=0$ {when} ${\partial}J_r/{\partial}J_m>0$, and
$\theta=1$ {when} ${\partial}J_r/{\partial}J_m<0$.
The thermal transistor is proper functioning under the condition $\beta_{l(r)}>1$.
Currently, it is known that the heat amplification can be realized mainly via two mechanisms:
(i) one is driven by the NDTC within the two-terminal setup,
where the heat current is suppressed with the increase of the temperature bias~\cite{bli2005apl,dhhe2014pre};
(ii) the other is driven by the inelastic transfer process without the NDTC, which can be unraveled even in the linear response regime~\cite{jhjiang2015prb}.


\begin{figure}[tbp]
\begin{center}
\includegraphics[scale=0.6]{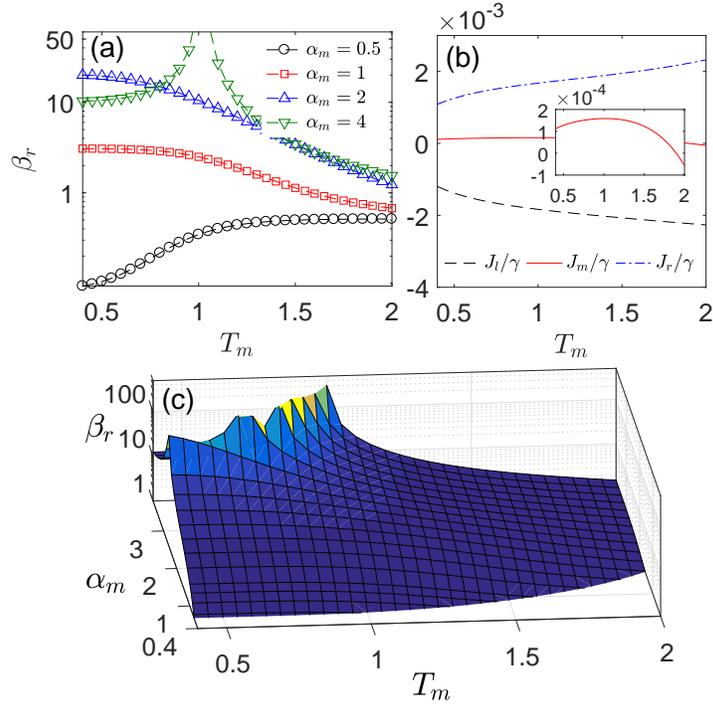}
\end{center}
\caption{(Color online) (a) Heat amplification factor $\beta_r$ as a function of the middle bath temperature $T_m$ with various system-middle bath coupling strength $\alpha_m$; (b) three steady state heat currents $J_u/\gamma~(u=l,m,r)$ as a function of $T_m$ with the coupling strength $\alpha_m=4$,
and the inset is the zoom in view of $J_m/\gamma$;
(c) the 3D view of the heat amplification factor $\beta_r$ by tuning $T_m$ and $\alpha_m$.
The other parameters are given by
$\varepsilon_l=1.0$, $\varepsilon_r=0.6$, $\Delta=0.6$, $\gamma=0.0002$, $\omega_c=10$,  $T_l=2$, and $T_r=0.4$.
}~\label{fig2p}
\end{figure}

\subsection{Transistor effect at strong coupling}

\subsubsection{Heat amplification}
We first investigate the influence of the strong system-middle bath interaction {on} the heat amplification by tuning the temperature $T_m$ of the middle bath.
As shown in Fig.~\ref{fig2p}(a), {the amplification factor is monotonically enhanced when the coupling strength increases from the moderate coupling regime (e.g., $\alpha_m=0.5$)}.
When the interaction strength enters the strong coupling regime ($\alpha_m=2$),
the amplification factor becomes large but finite in the low temperature regime of $T_m$ (see Appendix A2 for brief analysis), whereas it is strongly suppressed as $T_m$ reaches $T_m=T_l=2$.
Interestingly, as $\alpha_m$ is further strengthened (e.g., up to $4$), a giant heat amplification appears with a divergent point,
which results form the turnover behavior of $J_m$ shown in the inset of Fig.~\ref{fig2p}(b).
Moreover, the heat currents into the left and right thermal baths in Fig.~\ref{fig2p}(b) corresponding to $\alpha_m=4$ are much larger than $J_m$, which ensures the validity of the heat amplification in the strong coupling regime.

Next, we give a comprehensive picture of the amplification factor by modulating the temperature $T_m$
and coupling strength $\alpha_m$ in Fig.~\ref{fig2p}(c). It is found that the divergent behavior of the heat amplification is generally robust in the strong coupling regime ($\alpha_m{\gtrsim}2.8$). {In summary}, we conclude that the giant heat amplification feature favors the strong system-middle bath interaction.

\begin{figure}[tbp]
\begin{center}
\includegraphics[scale=0.4]{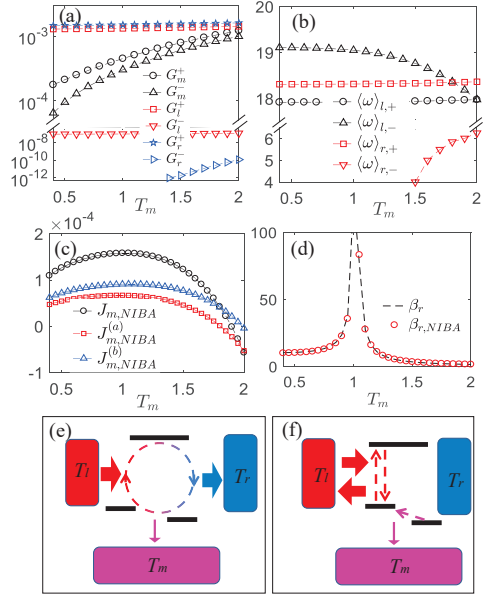}
\end{center}
\caption{(Color online) Steady state behaviors as a function the $T_m$ within the nonequilibrium NIBA at strong coupling ($\alpha_m=4$):
(a) transition rates $G^{\pm}_u~(u=l,m,r)$ at Eqs.~(\ref{gm}-\ref{gum}),
and (b) average energy quanta ${\langle}\omega{\rangle}_{u,\pm}~(u=l,r)$ at Eqs.~(\ref{oup}-\ref{oum});
(c) heat current $J_{m,\textrm{NIBA}}$ and its main components at Eqs.~(\ref{jm1}-\ref{jm2}),
and (d) comparison of the approximate amplification factor $\beta_{r,\textrm{NIBA}}$ with $\beta_r$;
(e) and (f) are schematic illustrations of flow components $J^{(a)}_{m,\textrm{NIBA}}$ and $J^{(b)}_{m,\textrm{NIBA}}$.
The other parameters are the same as in Fig.~\ref{fig2p}.
}~\label{fig3}
\end{figure}

\subsubsection{Mechanism of the giant heat amplification}
We devote this subsection to exploring the underlying mechanism of the giant heat amplification $\beta_r$
at strong system-middle bath coupling regime (e.g., $\alpha_m=4$) based on the analytical expression of heat currents in Eq.~(\ref{jl0}) and Eq.~(\ref{jr0}).
The limiting condition of large energy gap $(-E_r){\gg}1$ and low temperature of the right bath results in the vanishing of the average phonon number function $n_r(\omega{\approx}-E_r){\approx}0$.
Hence, the factor $G^-_{r}$ shows negligible contribution to the transition $|r{\rangle}\leftrightarrow|0{\rangle}$,
which is shown in Fig.~\ref{fig3}(a).
The large energy gap[$(-E_{l(r)}){\gg}1$] also generally leads to $G^+_{l(r)}{\gg}G^-_{l(r)}$.
Hence, the heat current {$J_r$} is simplified as
\begin{eqnarray}~\label{jrniba}
{J}_{r,\textrm{NIBA}}{\approx}\frac{1}{\mathcal{A}}G^-_lG^-_mG^+_r{\langle}\omega{\rangle}_{r,+},
\end{eqnarray}
with the coefficient $\mathcal{A}$ reduced to $\mathcal{A}{\approx}(G^+_m+G^-_m)(G^+_l+G^+_r)$.
${J}_{r,\textrm{NIBA}}$ is determined by the globally cyclic transition characterized by the rate $\frac{1}{\mathcal{A}}G^-_lG^-_mG^+_r$.
Moreover, it should be noted that though $G^+_m$ is much larger than $G^-_m$,
the ratio $G^-_m/G^+_m=\exp{[-(\varepsilon_l-\varepsilon_r)/(k_BT_m)]}$ shows monotonic increase as a function of $T_m$
[see dashed lines with circles and up-triangles in Fig.~\ref{fig3}(a)].
Then,
{by tuning up the temperature $T_m$ from $T_r=0.4$, the increase of $G^-_m/G^+_m$ dominates the monotonic enhancement of $J_{r,\textrm{NIBA}}$, as the rates $G^{\pm}_l$, $G^+_r$ and energy quanta ${\langle}\omega{\rangle}_{r,+}$ are nearly constant as shown in Fig.~\ref{fig3}(a) and Fig.~\ref{fig3}(b).

However, the current $J_{m,\textrm{NIBA}}$ is no longer a monotonic function of $T_m$,
which owns a maximum around $T_m=1$ as illustrated in Fig.~\ref{fig3}(e). The existence of a turnover point of $T_m$ is crucial to the giant heat amplification, so it  worths a careful study on $J_{m,\textrm{NIBA}}$. As $G^-_{r}$ is negligible, $J_{m,\textrm{NIBA}}$ can be approximated as the sum of two terms $J_{m,\textrm{NIBA}}{\approx}J^{(a)}_{m,\textrm{NIBA}}+J^{(b)}_{m,\textrm{NIBA}}$,
with components
\begin{subequations}
\begin{align}
J^{(a)}_{m,\textrm{NIBA}}{=}&\frac{1}{\mathcal{A}}G^-_mG^-_lG^+_r({\langle}\omega{\rangle}_{l,-}-{\langle}\omega{\rangle}_{r,+})~\label{jm1},\\~\label{jm2}
J^{(b)}_{m,\textrm{NIBA}}{=}&\frac{1}{\mathcal{A}}G^-_mG^-_lG^+_l({\langle}\omega{\rangle}_{l,-}-{\langle}\omega{\rangle}_{l,+}).
\end{align}
\end{subequations}
The approximate factor
$\beta_{r,\textrm{NIBA}}=\big{|}{\partial}J_{r,\textrm{NIBA}}/{\partial}(J^{(a)}_{m,\textrm{NIBA}}+J^{(b)}_{m,\textrm{NIBA}})\big{|}$
is agreeable with the counterpart obtained by the PRTE, shown in Fig.~\ref{fig3}(f).
Specifically, $J^{(a)}_{m,\textrm{NIBA}}$ describes a globally cyclic current with the loop rate ${G^-_mG^-_lG^+_r}/{\mathcal{A}}$ to extract energy quanta ${\langle}\omega{\rangle}_{l,-}$ out of the $l$th bath and
input ${\langle}\omega{\rangle}_{r,+}$ into the $r$th bath,
the resultant energy difference (${\langle}\omega{\rangle}_{l,-}-{\langle}\omega{\rangle}_{r,+}$) is absorbed by the middle bath.
While $J^{(b)}_{m,\textrm{NIBA}}$ is only associated with the local transition process between state $|l{\rangle}$ and $|0{\rangle}$, and each transition pumps energy quanta (${\langle}\omega{\rangle}_{l,-}-{\langle}\omega{\rangle}_{l,+}$) out of the left bath into the middle bath. These two currents are illustrated with Fig.~\ref{fig3}(c) and Fig.~\ref{fig3}(d), respectively.}

{For both $J^{(a)}_{m,\textrm{NIBA}}$ and $J^{(b)}_{m,\textrm{NIBA}}$, only the factors $G^-_m/G^+_m$ and $\langle\omega{\rangle}_{l,-}$ are obvious dependent on $T_m$, whereas all the other factors can be treated constant.
In the low temperature regime of $T_m$, the increase behavior of $J_{m,\textrm{NIBA}}$
is due to the increase of $G^-_m/G^+_m$. While as the temperature $T_m$ passing the turnover point, the monotonically decrease of ${\langle}\omega{\rangle}_{l,-}$ dominates the behavior of  $J_{m,\textrm{NIBA}}$[see Fig.~\ref{fig3}(b)].
}

Moreover, inspired by the giant heat amplification, we can realize the negative differential thermal conductance in a two-terminal setup,
which can be reduced from the current three terminal-model by eliminating the $r$th bath (see Appendix B for details).
With strong system-middle bath coupling (e.g., $\alpha_m=4$), the steady state current is approximately expressed as
\begin{eqnarray}~\label{j2t}
J_{l-m}{=}\frac{1}{\mathcal{A}^{\prime}}G^-_mG^+_lG^-_l({\langle}\omega{\rangle}_{l,-}-{\langle}\omega{\rangle}_{l,+}),
\end{eqnarray}
with $\mathcal{A}^{\prime}=(G^+_m+G^-_m)G^+_l+G^-_mG^-_l$.
It should be noted that $G^{\pm}_{l(m)}$ and ${\langle}\omega{\rangle}_{l,\pm}$
{in this two-terminal case} have the identical expression as shown at Eqs.~(\ref{gm}-\ref{gum})
and Eqs.~(\ref{oup}-\ref{oum}) within the three-terminal setup, respectively.
Hence, these quantities show the same behavior as exhibited in Fig.~\ref{fig3}(a) and (b).
Then, $J_{l-m}$ behaves quite similar to $J^{(b)}_{m,\textrm{NIBA}}$ at Eq.~(\ref{jm2}).
To this end, the NDTC  is expected by the turnover behavior of $J_{l-m}$.

\begin{figure}[tbp]
\begin{center}
\includegraphics[scale=0.5]{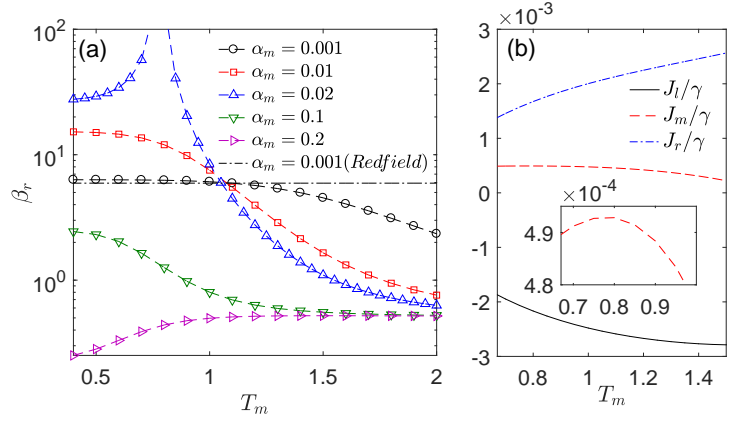}
\end{center}
\caption{(Color online) (a) Heat amplification factor $\beta_r$ with various coupling strengthes $\alpha_m$,
and (b) steady state heat currents with $\alpha_m=0.02$ as a function of $T_m$,
and the inset is the zoom-in view of $J_m/\gamma$.
The other parameters are the same as in Fig.~\ref{fig2p}.
}~\label{fig5}
\end{figure}

\subsection{Transistor effect at weak and moderate couplings}

\subsubsection{Heat amplification}
We investigate heat amplification at weak system-middle bath coupling in Fig.~\ref{fig5}(a).
It is found that in the weak coupling regime (e.g., $\alpha_m=0.001$ dashed black line with circle) the three-level system shows amplifying ability with finite amplification factor
($\beta_r{\approx}6$) in the low temperature regime $T_m{\in}[0.4,1.1]$.
This result is consistent with the counterpart from the Redfield equation (dashed-dotted line).
The phonon is not involved in the transition process between $|\pm{\rangle}_{\eta}$ and $|0{\rangle}$,
resulting in
$\textrm{Re}[\kappa_{u,+}(E_{\pm})]{\approx}\kappa^{(0)}_{u,+}(E_{\pm})={\kappa^{e}_{u,\pm}/2}$
and
$\textrm{Re}[\kappa_{u,-}(E_{\pm})]{\approx}\kappa^{(0)}_{u,-}(E_{\pm})={\kappa^{a}_{u,\pm}/2 }$.
The rates $\kappa^{e(a)}_{u,\pm}$ defined at Eqs.~(\ref{gep}-\ref{gem}).
Moreover, considering the limiting case $E_+{\gg}E_-$ and $\Gamma^{a(e)}_{+}{\gg}\Gamma^{a(e)}_-$,
the amplification factor is simplified as[see Eq.~(\ref{app-redfield-betar}) in Appendix B]
\begin{eqnarray}
\beta_r{\approx}\frac{\sin^2\theta}{16}\Big{|}\frac{\kappa^e_{l,+}\kappa^a_{r,+}-\kappa^a_{l,+}\kappa^e_{r,+}}
{\Gamma^a_-(\Gamma^a_++\Gamma^e_+)+\Gamma^a_+\Gamma^e_-}\Big{|}
\end{eqnarray}
which is irrelevant with the {$T_m$} and $\alpha_m$.

If we increase the coupling strength $\alpha_m$ up to the moderate regime (e.g., $\alpha_m=0.02$),
the giant amplification factor appears} in the comparatively low temperature regime [dashed line with up-triangle in Fig.~\ref{fig5}(a)],
which is due to the turnover behavior of $J_m$ in Fig.~\ref{fig5}(b).
Such feature results from the NDTC, which will be addressed in the {following subsection}.
It should be emphasized that the heat amplification is purely explored by the PTRE, which however cannot be explained with the Redfield equation.

\begin{figure}[tbp]
\begin{center}
\includegraphics[scale=0.35]{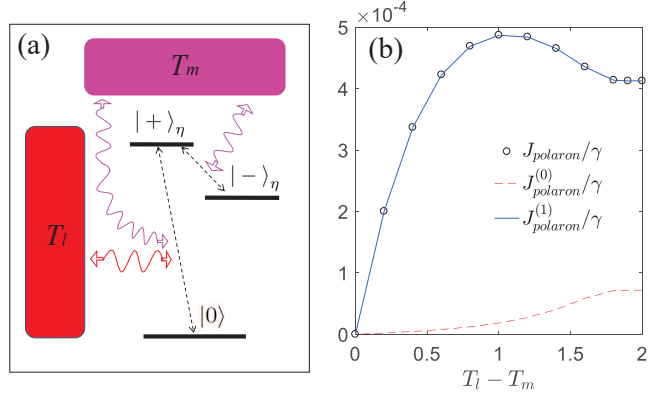}
\end{center}
\caption{(Color online) (a)Schematic illustration of quantum thermal transport in the three-level system (i.e. $|\pm{\rangle}_{\eta}$ and$|0{\rangle}$)
contacting with the $l$th and $m$th thermal baths with temperatures $T_l$ and $T_m$;
(b)steady state heat currents having different order approximations with the coupling strength $\alpha_m=0.02$.
The other parameters are the same as in Fig.~\ref{fig2p}.
}~\label{fig5-ndtc}
\end{figure}

\subsubsection{Negative differential thermal conductance}

{To better understand the NDTC,} we investigate the steady state heat current within the two-terminal setup (the $l$th and $m$th thermal baths) in Fig.~\ref{fig5-ndtc}(a).
Here, we stress that the phonon in the middle bath should be necessarily included to induce the NDTC.
Specifically, we keep one phonon transfer process for the rate $\gamma_{x(y)}(\omega)$.
While for rates $\kappa_{u,\pm}$, we first consider the zeroth order as
$\kappa^{(0)}_{u,+}(E_{\pm}){=}\eta^2_u\kappa^e_{u,\pm}/2$
and $\kappa^{(0)}_{u,-}(E_{\pm}){=}\eta^2_u\kappa^a_{u,\pm}/2$.
Then, the {zeroth order heat current $J^{(0)}$ obtained from the PTRE} shows monotonic enhancement by increasing the temperature bias $T_l-T_m$ in Fig.~\ref{fig5-ndtc}(b),
which demonstrates no NDTC signature.
Next, we include first order corrections to the transition rates as
\begin{subequations}
\begin{align}
\kappa^{(1)}_{u,+}(E_{\pm})=&\kappa^{(0)}_{u,+}(E_{\pm})+\int^\infty_{-\infty}\frac{d\omega_1}{4\pi}
\Lambda_u(\omega_1)n_u(\omega_1)\textrm{Re}[C^{(1)}_u(\omega_1-E_{\pm})],\\
\kappa^{(1)}_{u,-}(E_{\pm})=&\kappa^{(0)}_{u,-}(E_{\pm})+\int^\infty_{-\infty}\frac{d\omega_1}{4\pi}
\Lambda_u(\omega_1)[1+n_u(\omega_1)]\textrm{Re}[C^{(1)}_u(-\omega_1+E_{\pm})],
\end{align}
\end{subequations}
with the single phonon correlation function
\begin{eqnarray}
C^{(1)}_u(\pm\omega_1{\mp}E_{\pm})=\frac{\eta^2_u}{4}\int^\infty_0{d\tau}e^{i(\pm\omega_1{\mp}E_{\pm})\tau}\phi_m(\tau).
\end{eqnarray}
The heat current {$J^{(1)}$ up to the first order correction} shows interesting NDTC feature, which is almost identical with the exact numerical solution from the PTRE $J$.
Therefore, we conclude that the middle bath phonon induced transition between $|0{\rangle}$ and $|\pm{\rangle}_{\eta}$ is crucial
to the NDTC, as shown in Fig.~\ref{fig5-ndtc}(a), which cannot be found from the standard Redfield scheme.

\section{Conclusion}
To summarize, we study the steady state heat currents in the nonequilibrium three-level system interacting with three individual thermal baths.
We apply the PTRE combined with FCS to investigate the system density matrix and the resultant heat currents.
In the weak and strong system-middle bath coupling limits, we obtain the analytical expression of heat currents with the Redfield scheme
and nonequilibrium NIBA approach, which are consistent with the counterpart of the PTRE.
This extends the application of the PTRE to the nonequilibrium three-level models from the previous nonequilibrium (coupled) spin-boson models.

We also study the thermal transistor effect by tuning the system-middle bath coupling strength from weak to strong coupling regimes.
We first explore the giant heat amplification factor with strong coupling.
It is found that the globally cyclic current component and middle bath mediated local current component are crucial to exhibit the turnover behavior of the current into the middle bath.
The joint cooperation between the rates ratio $G^-_m/G^+_m$ assisted by the middle thermal bath and energy quanta ${\langle}\omega{\rangle}_{l,-}$ mainly results in
such heat amplification feature.
Next, we investigate heat amplification at weak and moderate system-middle bath couplings.
In the weak coupling regime, the finite heat amplification is discovered and analytically estimated by the Redfield scheme.
While in the moderate coupling regime, it is interesting to find another giant amplification signature,
which is mainly contributed by the middle bath assisted thermal transport between states $|\pm{\rangle}_{\eta}$ and $|0{\rangle}$.
Moreover, we also analyze the corresponding NDTC effect with the two-terminal setup.
It should be noted that such giant amplification and NDTC behaviors cannot be explained by the Redfield scheme, which clearly demonstrates
the wide application of the PTRE.

We hope the analysis of the heat amplification and negative differential thermal conductance may provide some theoretical insight in the design of
the quantum thermal transistor.

\section{Acknowledgements}
C.W. is supported by the National Natural Science Foundation of China under Grant No. 11704093.
D.Z.X. is supported by the National Natural Science Foundation of China under Grant No. 11705008 and Beijing Institute of Technology Research
Fund Program for Young Scholars.

\appendix

\section{Nonequilibrium NIBA scheme}

\subsection{Steady state heat currents}
In the strong system-middle bath coupling regime, the modified Hamiltonian {in} Eq.~(\ref{hps}) is reduced to
\begin{eqnarray}
\hat{H}^\prime_s=\overline{\varepsilon}\hat{N}+\delta\varepsilon\hat{\sigma}_z,
\end{eqnarray}
and the system-middle bath interaction {in} Eq.~(\ref{vprime1}) becomes
\begin{eqnarray}
V^\prime_p=e^{i\hat{B}}\hat{\sigma}_-+e^{-i\hat{B}}\hat{\sigma}_+.
\end{eqnarray}
Combining with  full counting statistics,
the dynamical equation of populations {in} Eq.~(\ref{dechi1}) is specified as
\begin{subequations}
\begin{align}
\frac{dP^\chi_l}{dt}=&-G_m(2\delta\varepsilon)P^\chi_l+G^{\chi_m}_m(-2\delta\varepsilon)P^\chi_r
-G_{l,-}(E_l)P^\chi_l+G^{\chi_l,\chi_m}_{l,+}(E_l)P^\chi_0\\
\frac{dP^\chi_r}{dt}=&-G_m(-2\delta\varepsilon)P^\chi_r+G^{\chi_m}_m(2\delta\varepsilon)P^\chi_l
-G_{r,-}(E_r)P^\chi_r+G^{\chi_r,\chi_m}_{r,+}(E_r)P^\chi_0\\
\frac{dP^\chi_0}{dt}=&-\sum_{u=l,r}G_{u,+}(E_u)P_0
+\sum_{u=l,r}G^{\chi_u,\chi_m}_{u,-}(E_u)P^\chi_u,
\end{align}
\end{subequations}
where the rates are
\begin{eqnarray}
G^{\chi_m}_m(\omega)=e^{i\omega\chi_m}\int^\infty_{-\infty}d{\tau}e^{i\omega\tau}\eta^2{e^{\phi_m(\tau)}},
\end{eqnarray}
and
\begin{subequations}
\begin{align}
G^{\chi_u,\chi_m}_{u,+}=&\frac{1}{4\pi}\int^\infty_{-\infty}{d\omega_1}{\Lambda}_u(\omega_1)n_u(\omega_1)e^{-i\omega_1\chi_u}e^{i(\omega_1-E_u)\chi_m}
[{C}_u(\omega_1-E_u)+H.c.],\\
G^{\chi_u,\chi_m}_{u,-}=&\frac{1}{4\pi}\int^\infty_{-\infty}{d\omega_1}{\Lambda}_u(\omega_1)[1+n_u(\omega_1)]e^{i\omega_1\chi_u}e^{i(E_u-\omega_1)\chi_m}
[{C}_u(E_u-\omega_1)+H.c.],
\end{align}
\end{subequations}
In absence of counting parameters, the steady state populations are obtained as
\begin{subequations}
\begin{align}
P_0=&\frac{1}{\mathcal{A}}(G^+_mG^-_r+G^-_mG^-_l+G^-_lG^-_r),\\
P_l=&\frac{1}{\mathcal{A}}(G^-_mG^+_l+G^-_mG^+_r+G^+_lG^-_r),\\
P_r=&\frac{1}{\mathcal{A}}(G^+_mG^+_l+G^+_mG^+_r+G^-_lG^+_r),
\end{align}
\end{subequations}
with the coefficient
$\mathcal{A}=(G^+_m+G^-_m)(G^+_l+G^+_r)+G^+_mG^-_r+G^-_mG^-_l+G^-_lG^+_r+G^-_r(G^+_l+G^-_l)$.
And the currents into the left and right baths are given by
\begin{eqnarray}
J_u=G^+_u{\langle}\omega{\rangle}_{u,+}P_0-G^-_u{\langle}\omega{\rangle}_{u,-}P_u,~(u=l,r)
\end{eqnarray}
with the energy quanta
\begin{subequations}
\begin{align}
{\langle}\omega{\rangle}_{u,+}=&\frac{1}{4{\pi}G^+_u}\int{d\omega_1}\omega_1{\Lambda}_u(\omega_1)[1+n_u(\omega_1)]
[{C}_u(-E_u-\omega_1)+H.c.],\\
{\langle}\omega{\rangle}_{u,-}=&\frac{1}{4{\pi}G^-_u}\int{d\omega_1}\omega_1{\Lambda}_u(\omega_1)n_u(\omega_1)
[{C}_u(\omega_1+E_u)+H.c.],
\end{align}
\end{subequations}
and
$E_u=\varepsilon_u-\sum_k|g_{k,m}|^2/\omega_k$.
Due to the conservation of heat energy $\sum_{u=l,m,r}J_u=0$, the current into the middle bath is given by
$J_m=-J_l-J_r$, which is specified as
\begin{eqnarray}
J_m&=&-\frac{1}{\mathcal{A}}[G^+_mG^+_lG^-_r({\langle}\omega{\rangle}_{l,+}-{\langle}\omega{\rangle}_{r,-})
+G^-_mG^-_lG^+_r({\langle}\omega{\rangle}_{r,+}-{\langle}\omega{\rangle}_{l,-})]\nonumber\\
&&-\frac{1}{\mathcal{A}}[(G^-_m+G^-_r)G^+_lG^-_l({\langle}\omega{\rangle}_{l,+}-{\langle}\omega{\rangle}_{l,-})
+(G^+_m+G^-_l)G^+_rG^-_r({\langle}\omega{\rangle}_{r,+}-{\langle}\omega{\rangle}_{r,-})]\nonumber\\
\end{eqnarray}

\begin{figure}[tbp]
\begin{center}
\includegraphics[scale=0.5]{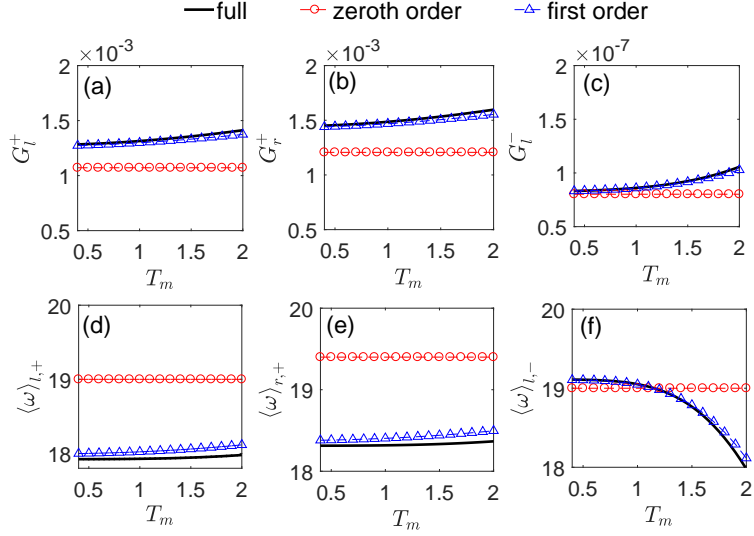}
\end{center}
\caption{(Color online)
The transition rate (a)$G^+_l$,(b)$G^+_r$,(c)$G^-_l$ and the energy quanta (d)${\langle}\omega{\rangle}_{l,+}$,
(e)${\langle}\omega{\rangle}_{r,+}$,(f)${\langle}\omega{\rangle}_{l,-}$ within the nonequilibrium NIBA scheme.
The solid black lines represent the full order calculation with expressions shown at Eqs.~(\ref{gup}-\ref{gum}) and Eqs.~(\ref{oup}-\ref{oum});
the dashed red lines with circles represent the zeroth order approximation at Eq.~(\ref{gup-zeroth}) and Eq.~(\ref{gml});
the dashed blue lines with up-triangles represent the first order approximation at Eq.~(\ref{gup-first-1}) and Eq.~(\ref{glm-first-1}).
The other parameters are given by
$\varepsilon_l=1.0$, $\varepsilon_r=0.6$, $\Delta=0.6$, $\gamma=0.0002$, $\omega_c=10$,  $T_l=2$, and $T_r=0.4$.
}~\label{app-order}
\end{figure}
\subsection{Influence of the middle bath in $G^{\pm}_u$ and ${\langle}{\omega}{\rangle}_{u,\pm}$ at strong qubit-middle bath coupling}
We first analyze the influence of the middle thermal bath in the rate $G^{+}_u$~($u=l,r$) and energy quanta ${\langle}{\omega}{\rangle}_{u,+}$.
It is known from Figs.~\ref{app-order}(a-b) and Figs.~\ref{app-order}(d-e) that {the middle bath} should be included to
properly describe $G^{+}_u$.
Specifically, the rate at Eq.~(\ref{gup}) with the first order correction is approximately expressed as
\begin{eqnarray}~\label{gup-first-1}
G^+_u&{\approx}&\frac{\eta^2_u}{2}\Lambda_u(-E_u)[1+n_u(-E_u)]\\
&&+\frac{\eta^2_u}{8\pi}\int^\infty_{-\infty}d\omega_1\frac{\Lambda_u(\omega_1)\Lambda_m(-E_u-\omega_1)}{(-E_u-\omega_1)^2}
[1+n_u(\omega_1)][1+n_m(-E_u-\omega_1)].\nonumber
\end{eqnarray}
In the low temperature regime of $T_m$, the Bose-Einstein distribution function $n_m(\omega>0)$ will be strongly suppressed
as $\omega$ deviates from zero, which may simplify $G^+_u$ to
\begin{eqnarray}~\label{gup-first-2}
G^+_u&{\approx}&\frac{\eta^2_u}{2}\Lambda_u(-E_u)[1+n_u(-E_u)]+\frac{\eta^2_u}{8\pi}\int^{E_u}_{-\infty}d\omega_1\frac{\Lambda_u(\omega_1)\Lambda_m(-E_u-\omega_1)}{(-E_u-\omega_1)^2}[1+n_u(\omega_1)],
\end{eqnarray}
which becomes insensitive to $T_m$.
Accordingly, the energy quanta can be obtained as
\begin{eqnarray}
{\langle}{\omega}{\rangle}_{u,+}=\frac{1}{4\pi{G^+_u}}\{2\pi{\eta^2_u}\Lambda_u(-E_u)[1+n_u(-E_u)](-E_u)
+\frac{1}{2}\int^{E_u}_{-\infty}d\omega_1\frac{\Lambda_u(\omega_1)\Lambda_m(-E_u-\omega_1)}{(-E_u-\omega_1)^2}
[1+n_u(\omega_1)]\}.
\end{eqnarray}
If we naively consider the zeroth order approximation, where the transitions are dominated by the resonant energy processes,
$G^+_u$ can be described as
\begin{eqnarray}~\label{gup-zeroth}
G^+_u{\approx}\frac{\eta^2_u}{2}\Lambda_u(-E_u)[1+n_u(-E_u)],
\end{eqnarray}
{and} ${\langle}{\omega}{\rangle}_{u,+}{\approx}-E_u$.

Next we study the behavior of $G^-_u$ by tuning $T_m$.
{Approximate to the first order of $\phi_m$},  $G^-_l$ can be expressed as
\begin{eqnarray}~\label{glm-first-1}
G^-_l&{\approx}&\frac{\eta^2_l}{2}\Lambda_u(-E_l)n_l(-E_l)\\
&&+\frac{\eta^2_l}{8\pi}\int^\infty_{-\infty}d\omega_1\frac{\Lambda_l(\omega_1)\Lambda_m(-E_l-\omega_1)}{(-E_l-\omega_1)^2}
n_l(\omega_1)n_m(-E_l-\omega_1),\nonumber
\end{eqnarray}
where the effect of one phonon needs to {be included} in the high temperature regime of $T_m$, as shown in Fig.~\ref{app-order}(c).
However, in the low temperature regime of $T_m$ (e.g.,$T_m{\approx}T_r$), $n_m(\omega>0)$ generally becomes negligible.
Moreover, the first order correction of one phonon from the middle bath contributes in the regime $\omega_1{\in}(-\infty,-E_u)$.
Hence, $G^-_l$ is dominated by the resonant process as
\begin{eqnarray}~\label{gml}
G^-_l&{\approx}&\frac{\eta^2_l}{2}\Lambda_l(-E_l)n_l(-E_l),
\end{eqnarray}
and thus ${\langle}\omega{\rangle}_{l,-}{\approx}-E_l$ [see Fig.~\ref{app-order}(f)].
{The leading order term of $G^-_r$ is $\frac{\eta^2_r}{2}\Lambda_r(-E_r)n_r(-E_r)$, which is negligible due to $n_r(-E_r){\approx}0$
in a wide regime of $T_m$.}

\begin{figure}[tbp]
\begin{center}
\includegraphics[scale=0.5]{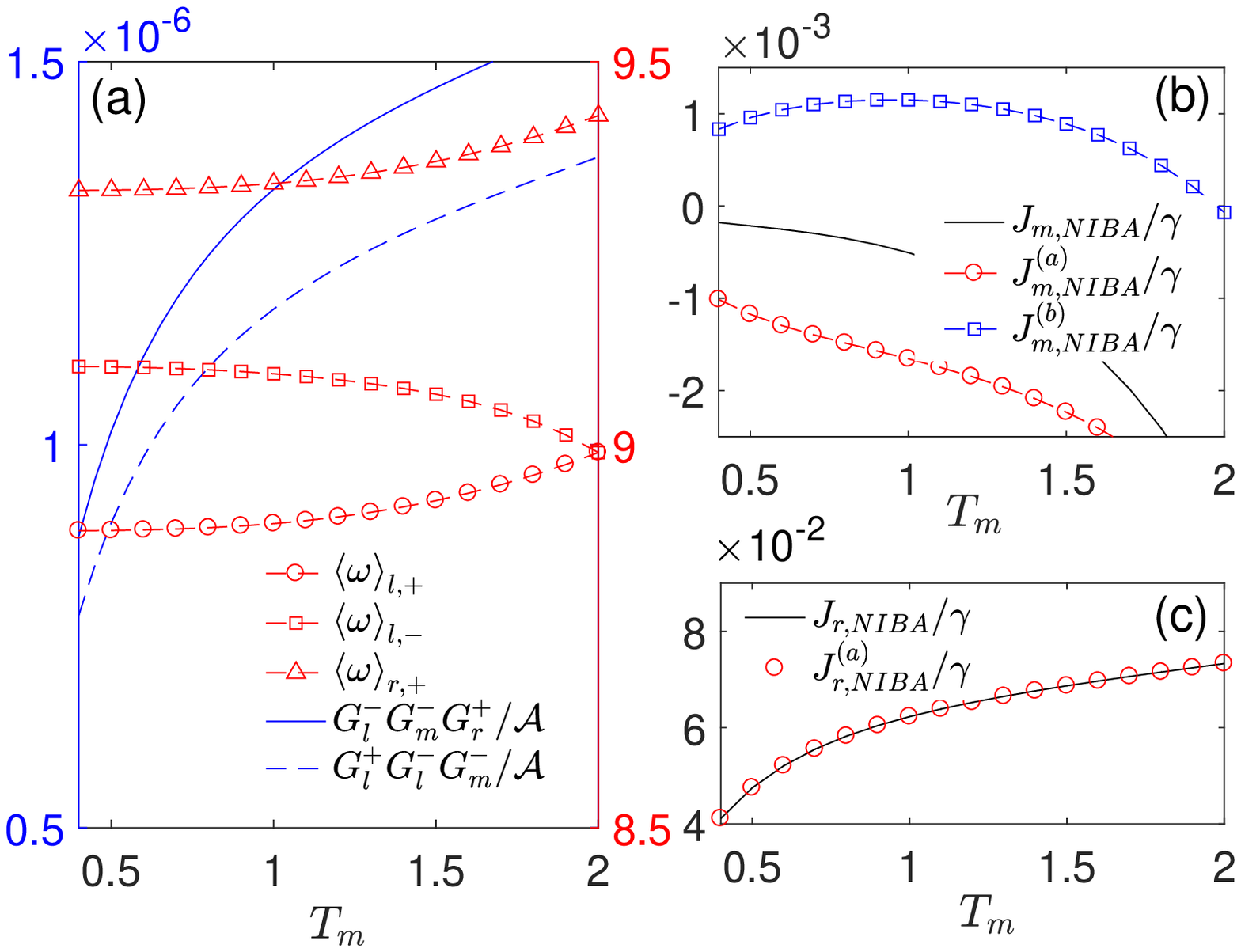}
\end{center}
\caption{(Color online) (a)Average energy quanta ${\langle}\omega{\rangle}_{l,\pm}$, ${\langle}\omega{\rangle}_{r,+}$,
and the flux rates $G^-_lG^-_mG^+_r/\mathcal{A}$ and $G^+_lG^-_lG^-_m/\mathcal{A}$;
(b)heat current into the middle bath $J_{m,\textrm{NIBA}}/\gamma$, the main components $J^{(1)}_{m,\textrm{NIBA}}/\gamma$
and $J^{(2)}_{m,\textrm{NIBA}}/\gamma$;
(c) heat current into the right bath $J_{r,\textrm{NIBA}}/\gamma$ and the main component
$J^{(1)}_{r,\textrm{NIBA}}/\gamma$
within the nonequilibrium NIBA scheme at strong coupling $\alpha_m=2$.
The other parameters are given by
$\varepsilon_l=1.0$, $\varepsilon_r=0.6$, $\Delta=0.6$, $\gamma=0.0002$, $\omega_c=10$,  $T_l=2$, and $T_r=0.4$.
}~\label{app-fig4}
\end{figure}
\subsection{Finite amplification factor at strong coupling}

At strong system-middle bath coupling $\alpha_m=2$, the current into the middle bath is approximated by
$J_{m,\textrm{NIBA}}{\approx}J^{(a)}_{m,\textrm{NIBA}}+J^{(b)}_{m,\textrm{NIBA}}$, with components
$J^{(a)}_{m,\textrm{NIBA}}=G^-_lG^-_mG^+_r({\langle}\omega{\rangle}_{l,-}-{\langle}\omega{\rangle}_{r,+})/\mathcal{A}$
and
$J^{(b)}_{m,\textrm{NIBA}}=G^-G^+_lG^-_l({\langle}\omega{\rangle}_{l,-}-{\langle}\omega{\rangle}_{l,+})/\mathcal{A}$,
as shown at Eq.~(\ref{jm1}) and Eq.~(\ref{jm2}).
From Fig.~\ref{app-fig4}(a), it is known that for $J^{(a)}_{m,\textrm{NIBA}}$
the magnitudes of both the flux rate $G^-_lG^-_mG^+_r/\mathcal{A}$ and the energy difference $({\langle}\omega{\rangle}_{l,-}-{\langle}\omega{\rangle}_{r,+})$ show monotonic increase,
which results in the enhancement of the  $J^{(a)}_{m,\textrm{NIBA}}$.
Moreover, $J^{(a)}_{m,\textrm{NIBA}}$ dominates the behavior of $J_{m,\textrm{NIBA}}$,
though $J^{(b)}_{m,\textrm{NIBA}}$ exhibits the turnover behavior, as shown in Fig.~\ref{app-fig4}(b).
While the heat current into the right bath is reduced to
$J^{(a)}_{r,\textrm{NIBA}}{\approx}\frac{G^-_lG^-_mG^+_r{\langle}\omega{\rangle}_{,r,+}}{\mathcal{A}}$,
as given at Eq.~(\ref{jrniba}).
And the flux rate $G^-_lG^-_mG^+_r/{\mathcal{A}}$ and the energy quanta ${\langle}\omega{\rangle}_{,r,+}$ are strengthened by increasing temperature $T_m$.
Hence, $J_{r,\textrm{NIBA}}$ shows the monotonic increase as in Fig.~\ref{app-fig4}(c).
considering the large deviation of magnitudes of currents $J_{r,\textrm{NIBA}}$ and $J_{m,\textrm{NIBA}}$,
the large and finite heat amplification $\beta_r$ is expected to observe.

\begin{figure}[tbp]
\begin{center}
\includegraphics[scale=0.5]{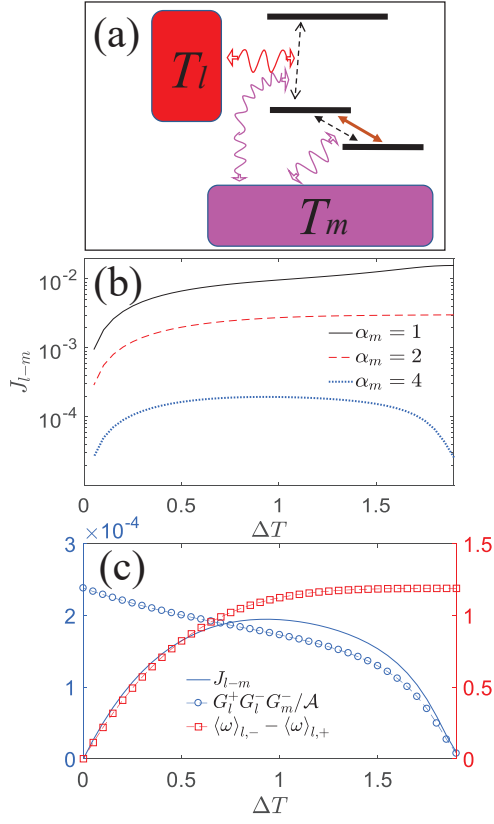}
\end{center}
\caption{(Color online) (a)Schematic illustration of quantum thermal transport in the three-level system (i.e. $|l(r){\rangle}_{\eta}$ and$|0{\rangle}$)
coupled to the $l$th and $m$th thermal baths with temperatures $T_l$ and $T_m$;
(b)steady state heat currents with different coupling strengthes from the PTRE;
(c)Heat current $J_{l-m}$, flux rates $G^+_lG^-_lG^-_m/\mathcal{A}$ and energy difference
$({\langle}\omega{\rangle}_{l,-}-{\langle}\omega{\rangle}_{l,+})$ at strong coupling $\alpha_m=4$ within the nonequilibrium NIBA.
The other parameters are the same as in Fig.~\ref{app-fig4}.
}~\label{app-fig4-2T}
\end{figure}
\subsection{Negative differential thermal conductance at strong coupling}

We investigate the steady state heat current in the three-level system {with a} two-terminal setup in Fig.~\ref{app-fig4-2T}(a) {by the PTRE}.
The nonmonotonic behavior of the current is clearly shown at strong system-middle bath coupling $\alpha_m=4$ in Fig.~\ref{app-fig4-2T}(b),
when the temperature bias ${\Delta}T=T_l-T_m$ increases.
{With certain approximation, the} NDTC can be explained within the nonequilibrium NIBA, {by which} the current is simplified to
$J_{l-m}=G^-_mG^+_lG^-_l({\langle}\omega{\rangle}_{l,-}-{\langle}\omega{\rangle}_{l,+})/\mathcal{A}$,
as {given in} Eq.~(\ref{j2t}).
Considering the monotonic suppression of the transition rate $G^-_m$ by increasing ${\Delta}T$ in Fig.~\ref{app-fig4-2T}(c),
i.e. by decreasing $T_m$ in Fig.~\ref{fig3}(a),
the transition from $|r{\rangle}$ to $|l{\rangle}$ is strongly blocked,
which suppresses the flux rate in $J_{l-m}$.
Therefore, the NDTC is clearly exhibited in Fig.~\ref{app-fig4-2T}(c).

\section{Redfield scheme}
The Hamiltonian of three-level system is given by
\begin{eqnarray}
\hat{H}_s=\sum_{u=l,r}\varepsilon_u|u{\rangle}{\langle}u|+\Delta(|l{\rangle}{\langle}r|+|r{\rangle}{\langle}l|)
\end{eqnarray}
{The} two excited eigenstates are given by
\begin{subequations}
\begin{align}
|+{\rangle}=&\cos\frac{\theta}{2}|l{\rangle}+\sin\frac{\theta}{2}|r{\rangle},\\
|-{\rangle}=&-\sin\frac{\theta}{2}|l{\rangle}+\cos\frac{\theta}{2}|r{\rangle},
\end{align}
\end{subequations}
with $\tan\theta=\Delta/\delta\varepsilon$, the eigenenergy
$E_{\pm}=\overline{\varepsilon}\pm\sqrt{\delta\varepsilon^2+\Delta^2}$,
$\overline{\varepsilon}=(\varepsilon_l+\varepsilon_r)/2$
and $\delta\varepsilon=(\varepsilon_l-\varepsilon_r)/2$.
The system-phonon interaction is given by
\begin{eqnarray}
\hat{V}_m=\sum_k(g_{k,m}\hat{b}^\dag_{k,m}+g^*_{k,m}\hat{b}_{k,m})\hat{S}_m
\end{eqnarray}
with $\hat{S}_m=\cos\theta(|+{\rangle}{\langle}+|-|-{\rangle}{\langle}-|)-\sin\theta(|+{\rangle}{\langle}-|+|-{\rangle}{\langle}+|)$.
Then, the dynamical equation is given by
\begin{eqnarray}
\frac{d\hat{\rho}_{\chi}}{dt}=-i[\hat{H}_s,\hat{\rho}_{\chi}]+\sum_u\mathcal{L}^{u}_{\chi_u}[\hat{\rho}_{\chi}],
\end{eqnarray}
where the dissipator related with the middle bath is
\begin{eqnarray}
\mathcal{L}^{\chi_m}_m[\hat{\rho}_{\chi}]&=&\sum_{\omega,\omega^\prime}\{
-{J_m(\omega^\prime)}n_m(\omega^\prime)[\hat{\rho}_{\chi}\hat{S}_m(\omega^\prime)\hat{S}_m(\omega)+\textrm{H.c.}]\\
&&+e^{-i\omega^\prime\chi_m}{J_m(\omega^\prime)}n_m(\omega^\prime)\hat{S}_m(\omega)\hat{\rho}_{\chi}\hat{S}_m(\omega^\prime)\nonumber\\
&&+e^{i\omega^\prime\chi_m}{J_m(\omega^\prime)}[1+n_m(\omega^\prime)]\hat{S}_m(\omega^\prime)\hat{\rho}_{\chi}\hat{S}_m(\omega)\},\nonumber
\end{eqnarray}
with $\hat{S}_m(-\tau)=\sum_{\omega}\hat{S}_m(\omega)e^{i\omega\tau}$.
And the dissipator related with the $l(r)$th bath is
\begin{eqnarray}
\mathcal{L}^u_{\chi_u}[\hat{\rho}_{\{\chi\}}]&=&
\frac{J_u(\omega^\prime)}{4}n_u(\omega^\prime)e^{-i\omega^\prime\chi_u}[\hat{Q}^\dag_u(\omega^\prime)\hat{\rho}_{\{\chi\}}\hat{Q}_u(\omega)+\hat{Q}^\dag_u(\omega)\hat{\rho}_{\{\chi\}}\hat{Q}_u(\omega^\prime)]\\
&&+\frac{J_u(\omega^\prime)}{4}(1+n_u(\omega^\prime))e^{i\omega^\prime\chi_u}[\hat{Q}_u(\omega^\prime)\hat{\rho}_{\{\chi\}}\hat{Q}^\dag_u(\omega)+\hat{Q}_u(\omega)\hat{\rho}_{\{\chi\}}\hat{Q}^\dag_u(\omega^\prime)]\nonumber\\
&&-[\frac{J_u(\omega^\prime)}{4}n_u(\omega^\prime)\hat{Q}_u(\omega)\hat{Q}^\dag_u(\omega^\prime)\hat{\rho}_{\{\chi\}}
+\frac{J_u(\omega^\prime)}{4}(1+n_u(\omega^\prime))\hat{Q}^\dag_u(\omega)\hat{Q}_u(\omega^\prime)\hat{\rho}_{\{\chi\}}+\textrm{H.c.}]\nonumber
\end{eqnarray}
with
$\hat{S}_u(-\tau)=\sum_{\omega}\hat{Q}_u(\omega)e^{i\omega\tau}$.

The steady state heat current obtained by  FCS is given by
\begin{eqnarray}
J_l&=&\frac{{\cos^2\frac{\theta}{2}}}{2}\kappa^a_{l,+}E_+\rho_{++}+\frac{\sin^2\frac{\theta}{2}}{2}\kappa^a_{l,-}E_-\rho_{--}
-(\frac{\cos^2\frac{\theta}{2}}{2}\kappa^e_{l,+}E_++\frac{\sin^2\frac{\theta}{2}}{2}\kappa^e_{l,-}E_-)\rho_{00},\\
J_r&=&\frac{\sin^2\frac{\theta}{2}}{2}\kappa^a_{r,+}E_+\rho_{++}+\frac{\cos^2\frac{\theta}{2}}{2}\kappa^a_{r,-}E_-\rho_{--}
-(\frac{\sin^2\frac{\theta}{2}}{2}\kappa^e_{r,+}E_++\frac{\cos^2\frac{\theta}{2}}{2}\kappa^e_{r,-}E_-)\rho_{00},\\
J_m&=&\frac{\sin^2\theta}{2}(E_+-E_-)(\kappa^a_m\rho_{++}-\kappa^e_m\rho_{--}),
\end{eqnarray}
and
$J_m=-J_l-J_r$. It should be noted that the steady state coherence in the eigenbasis is negligible.

Moreover, the steady state populations are given by
\begin{subequations}
\begin{align}
P_+=&\frac{1}{\mathcal{B}}(\Gamma^e_++\Gamma^e_-)[\Gamma^e_+\Gamma^a_-+(\Gamma^e_++\Gamma^e_-)\Gamma^+_p],\\
P_-=&\frac{1}{\mathcal{B}}(\Gamma^e_++\Gamma^e_-)[\Gamma^a_+\Gamma^e_-+(\Gamma^e_++\Gamma^e_-)\Gamma^-_p],\\
P_0=&\frac{1}{\mathcal{B}}(\Gamma^a_+[\Gamma^e_+\Gamma^a_-+(\Gamma^e_++\Gamma^e_-)\Gamma^+_p]
+\Gamma^a_-[\Gamma^a_+\Gamma^e_-+(\Gamma^e_++\Gamma^e_-)\Gamma^-_p]),\\
\mathcal{B}=&(\Gamma^a_++\Gamma^e_++\Gamma^e_-)[\Gamma^e_+\Gamma^a_-+(\Gamma^e_++\Gamma^e_-)\Gamma^+_p]
+(\Gamma^a_-+\Gamma^e_++\Gamma^e_-)[\Gamma^a_+\Gamma^e_-+(\Gamma^e_++\Gamma^e_-)\Gamma^-_p],
\end{align}
\end{subequations}
with the rates defined as
\begin{subequations}
\begin{align}
\Gamma^{e(a)}_+=&\frac{1}{2}(\kappa^{e(a)}_{l,+}\cos^2\frac{\theta}{2}+\kappa^{e(a)}_{r,+}\sin^2\frac{\theta}{2}),\\
\Gamma^{e(a)}_-=&\frac{1}{2}(\kappa^{e(a)}_{l,-}\sin^2\frac{\theta}{2}+\kappa^{e(a)}_{r,-}\cos^2\frac{\theta}{2}),\\
\Gamma^{+(-)}_p=&\frac{\sin^2\theta}{8}\kappa^{e(a)}_p,
\end{align}
\end{subequations}
and
$\kappa^e_{u,\pm}=\Lambda_u(E_{\pm})n_u(E_{\pm})$,
$\kappa^a_{u,\pm}=\Lambda_u(E_{\pm})[1+n_u(E_{\pm})]$,
$\kappa^e_{p}=\Lambda_m(E_{+}-E_-)n_m(E_{+}-E_-)$,
and
$\kappa^e_{p}=\Lambda_m(E_{+}-E_-)[1+n_m(E_{+}-E_-)]$.
The currents are specified as
\begin{eqnarray}
J_l&=&\sum_{\xi=\pm}\frac{(1+\xi\cos\theta)}{4\mathcal{B}}E_{\xi}[\kappa^a_{l,\xi}(\Gamma^e_++\Gamma^e_-)\Gamma^e_{\xi}\Gamma^a_{\bar{\xi}}
-\kappa^e_{l,\xi}(\Gamma^a_+\Gamma^e_+\Gamma^a_-+\Gamma^a_-\Gamma^e_-\Gamma^a_+)]\nonumber\\
&&+\sum_{\xi=\pm}\frac{(1+\xi\cos\theta)}{4\mathcal{B}}E_{\xi}(\Gamma^e_++\Gamma^e_-)
[\kappa^a_{l,\xi}(\Gamma^e_++\Gamma^e_-)\Gamma^{\xi}_p-\kappa^e_{l,\xi}(\Gamma^a_+\Gamma^+_p+\Gamma^a_-\Gamma^-_p)],\\
J_r&=&\sum_{\xi=\pm}\frac{(1-\xi\cos\theta)}{4\mathcal{B}}E_{\xi}[\kappa^a_{r,\xi}(\Gamma^e_++\Gamma^e_-)\Gamma^e_{\xi}\Gamma^a_{\bar{\xi}}
-\kappa^e_{r,\xi}(\Gamma^a_+\Gamma^e_+\Gamma^a_-+\Gamma^a_-\Gamma^e_-\Gamma^a_+)]\nonumber\\
&&+\sum_{\xi=\pm}\frac{(1-\xi\cos\theta)}{4\mathcal{B}}E_{\xi}(\Gamma^e_++\Gamma^e_-)
[\kappa^a_{r,\xi}(\Gamma^e_++\Gamma^e_-)\Gamma^{\sigma}_p-\kappa^e_{r,\sigma}(\Gamma^a_+\Gamma^+_p+\Gamma^a_-\Gamma^-_p)].
\end{eqnarray}
The current into the middle bath is
\begin{eqnarray}
J_m=-\frac{1}{\mathcal{B}}(E_+-E_-)(\Gamma^e_++\Gamma^e_-)(\Gamma^a_+\Gamma^e_-\Gamma^+_p-\Gamma^a_-\Gamma^e_+\Gamma^-_p).
\end{eqnarray}

In particular, under the condition $E_+{\gg}E_-$ and $\Gamma^{\pm}_p{\gg}\Gamma^{a(e)}_-$,
the coefficient $\mathcal{B}$ is reduced to
$\mathcal{B}=(\Gamma^e_++\Gamma^-_e)[(\Gamma^a_++\Gamma^e_++\Gamma^e_-)\Gamma^+_p+(\Gamma^a_-\Gamma^e_++\Gamma^e_-)\Gamma^-_p]$,
and currents into the middle and right baths are approximated as
\begin{eqnarray}
J_m&{\approx}&-\frac{E_+(\Gamma^e_++\Gamma^e_-)}{\mathcal{B}}(\Gamma^a_+\Gamma^e_-\Gamma^+_p-\Gamma^a_-\Gamma^e_+\Gamma^-_p),\\
J_r&{\approx}&\frac{(1-\cos\theta)E_+(\Gamma^e_++\Gamma^e_-)}{4\mathcal{B}}
[\kappa^a_{r,+}(\Gamma^e_++\Gamma^e_-)\Gamma^+_p-\kappa^e_{r,+}(\Gamma^a_+\Gamma^+_p+\Gamma^a_-\Gamma^-_p)].
\end{eqnarray}
If we redefine $J_m=J^{\prime}_m+J^0_m$ and $J_r=J^{\prime}_r+J^0_r$,
with
\begin{eqnarray}
J^{\prime}_m=E_+{\times}\frac{\Gamma^a_-\Gamma^e_+}{(\Gamma^a_-+\Gamma^e_++\Gamma^e_-)}
{\times}\frac{(\Gamma^a_+\Gamma^e_-)/(\Gamma^a_-\Gamma^e_+)+(\Gamma^a_++\Gamma^e_++\Gamma^e_-)/(\Gamma^a_-+\Gamma^e_++\Gamma^e_-)}
{\Gamma^-_p/\Gamma^+_p+(\Gamma^a_++\Gamma^e_++\Gamma^e_-)/(\Gamma^a_-+\Gamma^e_++\Gamma^e_-)},
\end{eqnarray}
$J^0_m=E_+\Gamma^a_-\Gamma^e_+/(\Gamma^a_-+\Gamma^e_++\Gamma^e_-)$,
\begin{eqnarray}
J^{\prime}_r&=&\frac{(1-\cos\theta)E_+}{4}{\times}\frac{\kappa^e_{r,+}\Gamma^a_-}{(\Gamma^a_-+\Gamma^e_++\Gamma^e_-)}\\
&&{\times}
\frac{(\Gamma^a_++\Gamma^e_++\Gamma^e_-)/(\Gamma^a_-+\Gamma^e_++\Gamma^e_-)+[\kappa^a_{r,+}(\Gamma^e_++\Gamma^e_-)-\kappa^e_{r,+}\Gamma^a_+]
/(\kappa^e_{r,+}\Gamma^a_-)}{\Gamma^-_p/\Gamma^+_p+(\Gamma^a_++\Gamma^e_++\Gamma^e_-)/(\Gamma^a_-+\Gamma^e_++\Gamma^e_-)},\nonumber
\end{eqnarray}
and
$J^0_r=-\frac{(1-\cos\theta)E_+}{4}\kappa^e_{r,+}\Gamma^a_-/(\Gamma^a_-+\Gamma^e_++\Gamma^e_-)$.
It should be noted that $J^0_m$ and $J^0_r$ are irrelevant with $T_m$.
Hence, the linear heat amplification is given by
\begin{eqnarray}
\beta_r{\approx}\frac{(1-\cos\theta)}{4}\Big{|}\frac{\kappa^e_{r,+}\Gamma^a_-(\Gamma^a_++\Gamma^e_++\Gamma^e_-)+(\Gamma^a_-+\Gamma^e_++\Gamma^e_-)
[\kappa^a_{r,+}(\Gamma^e_++\Gamma^e_-)-\kappa^e_{r,+}\Gamma^a_{+}]}{\Gamma^a_-\Gamma^e_+(\Gamma^a_++\Gamma^e_++\Gamma^e_-)
+(\Gamma^a_-+\Gamma^e_++\Gamma^e_-)\Gamma^a_+\Gamma^e_-}\Big{|}.
\end{eqnarray}
Moreover, considering the condition $\Gamma^{a(e)}_{+}{\gg}\Gamma^{a(e)}_-$,
Hence, the amplification factor is simplified as
\begin{eqnarray}~\label{app-redfield-betar}
\beta_r{\approx}\frac{\sin^2\theta}{16}\Big{|}\frac{\kappa^e_{l,+}\kappa^a_{r,+}-\kappa^a_{l,+}\kappa^e_{r,+}}
{\Gamma^a_-(\Gamma^a_++\Gamma^e_+)+\Gamma^a_+\Gamma^e_-}\Big{|}
\end{eqnarray}

\end{document}